\documentclass[lettersize,journal]{IEEEtran}
\IEEEoverridecommandlockouts
\usepackage{graphicx, amsmath, amsthm, amssymb, subcaption, url, cite, array, amsthm,booktabs,xcolor}
\usepackage[ruled, linesnumbered]{algorithm2e} 
\usepackage{algorithm2e, setspace}
\usepackage[font={small}]{caption}
\usepackage{hyperref}
\usepackage{fancyhdr}

\usepackage{lipsum}  
\IEEEoverridecommandlockouts
\makeatletter
\let\@oldmaketitle\@maketitle
\renewcommand{\@maketitle}{
  \@oldmaketitle
  \vspace{-10pt}  
}
\makeatother

\begin{document}
\title{\huge Energy-Efficient and Intelligent ISAC in V2X Networks with Spiking Neural Networks-Driven DRL
}

\author{Chen Shang, Jiadong Yu,~\IEEEmembership{Member,~IEEE,}
Dinh Thai Hoang,~\IEEEmembership{Senior Member,~IEEE}
% \vspace{-5pt}
\thanks{This work is supported by Guangdong Provincial Key Lab of Integrated Communication, Sensing and Computation for Ubiquitous Internet of Things (No.2023B1212010007). (\textit{Corresponding author: Jiadong Yu.})

Chen Shang and Jiadong Yu are with the Internet of Things Thrust, The Hong Kong University
of Science and Technology (Guangzhou), Guangzhou, Guangdong, China (chenshang@hkust-gz.edu.cn, jiadongyu@hkust-gz.edu.cn). Dinh Thai Hoang is with the School of Electrical and Data Engineering, University of Technology Sydney, Australia (hoang.dinh@uts.edu.au).}}
\maketitle
% \maketitle
 \thispagestyle{fancy}
\pagestyle{fancy}
\lhead{This paper has been accepted by IEEE Transactions on Wireless Communications.}
\rhead{}
\cfoot{\thepage}
\renewcommand{\headrulewidth}{0.4pt}
\renewcommand{\footrulewidth}{0pt}

% DOI: \href{https://doi.org/10.1109/TWC.2023.3326884}{https://doi.org/10.1109/TWC.2023.3326884} 

\begin{abstract}
Integrated sensing and communication (ISAC) is emerging as a key enabler for vehicle-to-everything (V2X) systems. However, designing efficient beamforming schemes for ISAC signals to achieve accurate sensing and enhance communication performance in the dynamic and uncertain environments of V2X networks presents significant challenges. While artificial intelligence technologies offer promising solutions, the energy-intensive nature of neural networks imposes substantial burdens on communication infrastructures. To address these challenges, this work proposes an energy-efficient and intelligent ISAC system for V2X networks. Specifically, we first leverage a Markov Decision Process framework to model the dynamic and uncertain nature of V2X networks. This framework allows the roadside unit to develop beamforming schemes relying solely on its current sensing information, eliminating the need for numerous pilot signals and extensive CSI acquisition. We then introduce an advanced deep reinforcement learning (DRL) algorithm, enabling the joint optimization of beamforming and power allocation to guarantee both communication rate and sensing accuracy in dynamic and uncertain V2X scenario. To alleviate the energy demands of neural networks, we integrate spiking neural networks (SNNs) into the DRL algorithm. The event-driven, sparse spike-based processing of SNNs significantly improves energy efficiency while maintaining strong performance. Extensive simulation results validate the effectiveness of the proposed scheme with lower energy consumption, superior communication performance, and improved sensing accuracy.
\end{abstract}

% \vspace{-10pt}
\section{Introduction}
The rapid advancement of wireless communication technologies has become the backbone of modern autonomous vehicles and intelligent transportation systems, driving unprecedented demands for performance and reliability~\cite{10529954}. These systems rely on ultra-high data rates to enable bandwidth-intensive applications like high-definition mapping, cooperative driving, and real-time video streaming. At the same time, they require ultra-low latency for instantaneous responses in safety-critical scenarios, such as collision avoidance and emergency braking, alongside massive connectivity to ensure seamless interaction among vehicles, roadside units (RSUs), and cloud infrastructure, i.e., vehicle-to-everything (V2X) networks.
To meet these escalating demands, massive multiple-input multiple-output (MIMO) and millimeter-wave (mmWave) technologies have emerged as transformative solutions~\cite{8241348}. Massive MIMO enhances spectral efficiency by leveraging large antenna arrays to serve multiple users simultaneously via spatial multiplexing, making it ideal for dense, dynamic V2X networks. Meanwhile, mmWave exploits abundant spectrum in millimeter-wave bands to deliver unparalleled data rates and capacity. Together, these technologies enable ultra-fast communication rates and reliable performance, establishing their indispensability in V2X networks~\cite{10529954,8241348}.

Despite their advantages, deploying these technologies to enhance V2X networks faces significant challenges. First, the high data rates provided by MIMO heavily depend on beam training, which involves a multi-step process to achieve precise beam alignment. For example, the RSU initially transmits pilot signals to the vehicle, which uses these signals to estimate the channel state information (CSI) and feeds this information back to the RSU. During this process, both the RSU and the vehicle periodically transmit and receive pilot signals across all potential beam directions to identify the beam pair that maximizes path gain. Based on this feedback, the RSU dynamically adjusts its beam direction to align with the vehicle’s position and movement. While this iterative process ensures optimal signal strength and coverage, it also requires substantial pilot signals for channel estimation and beam alignment. As a result, it significantly increases resource consumption, reduces spectral efficiency, and introduces overhead delays that are especially critical in the fast-changing environments of V2X networks~\cite{10634048}. To address this issue, endowing the V2X system with sensing capabilities through radar deployment has emerged as a potential solution to reduce beam training overhead.
However, conventional approaches that separately utilize radar signals for sensing and communication signals for data transmission further exacerbate spectral resource inefficiency~\cite{10529954}. Therefore, these combined requirements for robust communication and accurate sensing expose significant limitations of conventional wireless technologies in meeting the demands of modern V2X networks.

The paradigm shift from 5G to 6G networks introduces transformative advancements in wireless communication through integrated sensing and communication (ISAC)~\cite{10286339} and artificial intelligence (AI)~\cite{10663823}. By utilizing sensing signal to estimate vehicle's state information, ISAC reduces reliance on resource-intensive beam training, enhances beamforming efficiency, and optimizes spectrum utilization~\cite{9737357}. Meanwhile, AI-driven wireless communication introduces intelligent beam management, dynamic resource allocation, and adaptive learning capabilities, enabling real-time adaptation to rapidly changing V2X environments. Together, ISAC and AI synergize to enhance beam alignment precision, strengthen communication robustness, and reduce resource overhead, paving the way for reliable V2X connectivity, mobility, and security. 

% The paradigm shift from 5G to 6G networks introduces transformative advancements in wireless communication through integrated sensing and communication (ISAC) and artificial intelligence (AI)~\cite{10663823}. Specifically, the dual-functional radar-communication (DFRC) system of ISAC integrates communication and sensing into a unified framework, allowing shared resources to perform both of communication and sensing tasks simultaneously~\cite{9737357}. By utilizing sensing signal to predict vehicle's state information, ISAC reduces reliance on resource-intensive beam training, enhances beamforming efficiency, and optimizes spectrum utilization~\cite{9737357}. Meanwhile, AI-driven wireless communication introduces intelligent beam management, dynamic resource allocation, and adaptive learning capabilities, enabling real-time adaptation to rapidly changing V2X environments. Together, ISAC and AI synergize to enhance beam alignment precision, strengthen communication robustness, and reduce resource overhead, paving the way for reliable V2X connectivity, mobility, and security. 

% \vspace{-16.5pt}
\subsection{Related Work}
% \vspace{-4pt}
Within ISAC-assisted V2X networks, the primary objective is to develop efficient beamforming strategies that enhance sensing accuracy and subsequently optimize communication performance, i.e., designing beamforming for sensing-assisted communication. Currently, beamforming schemes can be realized through optimization-based approaches \cite{liu2020radar,10659350,10561505,10600135} and AI-driven techniques~\cite{liu2022learning,10605608,10571018,10304580}.

In~\cite{liu2020radar}, an Extended Kalman Filter (EKF) was employed to track the motion of vehicles using sensing signals, facilitating the update of beamforming parameters to ensure consistent communication performance. A similar method was adopted in~\cite{10659350}, where the EKF was extended to track both vehicle and unmanned aerial vehicle (UAV) kinematics. This enabled the UAV to dynamically adjust beam widths and trajectories to maintain robust sensing and communication in high-mobility environments. Additionally, \cite{10561505} introduced a predictive beam tracking approach that integrated vehicle-sensed data with sensing information to enhance trajectory prediction accuracy, mitigate multi-user interference, and improve communication consistency in nonlinear V2X scenarios. To address inter-beam and inter-vehicle interference, \cite{10600135} proposed a dynamic power allocation strategy that optimized the trade-offs between sensing and communication.

AI-driven beamforming schemes, on the other hand, leverage advanced learning algorithms to optimize beamforming strategies. In~\cite{liu2022learning}, a hybrid neural network framework combining convolutional neural networks (CNNs) and a long short-term memory (LSTM) network was proposed to predict the current beamforming matrix based on historical channel data. This approach eliminated the need for explicit channel tracking, significantly reducing signaling overhead. Besides,\cite{10605608} developed three neural networks to independently handle communication, sensing, and power allocation tasks, achieving greater system efficiency. In addition,\cite{10571018} utilized deep reinforcement learning (DRL) to jointly optimize beamforming and power allocation in static V2X networks, offering a model-free approach to complex optimization tasks. Furthermore,~\cite{10304580} introduced a two-stage predictive beamforming framework employing deep neural networks. The first stage acquired CSI, while the second stage optimized the beamforming matrices to enhance communication performance.

Although the aforementioned methods have significantly enhanced the performance of ISAC systems, unresolved challenges remain. First, conventional optimization-based approaches~\cite{liu2020radar,10659350,10561505} that rely on the EKF to predict vehicle states and perform beamforming. However, these approaches can encounter challenges in modeling nonlinear vehicle dynamics accurately. For instance, the EKF relies on linearization via first-order Taylor expansion, which may lead to approximation errors that accumulate over time, especially in highly dynamic V2X scenarios. Although advanced techniques such as the Unscented Kalman Filter (UKF)~\cite{xiong2006performance} provide better handling of nonlinearities, they often assume known and stationary process and measurement noise statistics. These assumptions may not hold in practical ISAC-assisted V2X environments, where sensor uncertainties and communication-induced delays can be time-varying and difficult to model precisely.

On the other hand, AI-based schemes~\cite{liu2022learning,10605608,10571018,10304580} utilized data-driven methods to estimate and predict CSI, bypassing the need for explicit motion tracking and directly enabling beamforming design. Although effective in some cases, these methods face difficulties in highly dynamic V2X environments, where rapid channel variations and interference undermine CSI reliability. 

A critical limitation of AI-driven algorithms are inherently energy-intensive, posing substantial energy consumption challenges during training, inference, and model updates. This high energy demand arises from the computational complexity of forward and backward propagations in deep neural networks, particularly in architectures with a large number of parameters and layers~\cite{liu2022learning,10605608,10571018,10304580,10229017}. During training, these algorithms require iterative optimization over massive datasets, significantly increasing energy usage. For example, training a large-scale neural network like GPT-3, which contains 175 billion parameters, consumed approximately 1,287 megawatt-hours (MWh) of electricity-comparable to the energy usage of 130 U.S. homes over a year~\cite{de2023growing}. The computational burden is further amplified during the practical deployment of these algorithms, which requires the RSU to continuously perform real-time, sustainable processing in order to adapt to fluctuating channel conditions and dynamic vehicle states. Moreover, maintaining model performance in non-stationary environments necessitates frequent updates, imposing continuous computational and energy demands.
Notably, Google has reported that 60$\%$ of its AI-related energy consumption from 2019 to 2021 was attributed to inference processes~\cite{de2023growing}.
These challenges collectively place substantial pressure on energy-limited communication infrastructures, e.g., RSUs. The high energy consumption associated with training, inference, and model updates not only limits the scalability of AI-driven methods but also threatens the feasibility of their deployment in practical V2X scenarios.

% \vspace{-6pt}
\subsection{Motivation and Contributions}
% \vspace{-5pt}
Given the above challenges, this work aims to develop an intelligent and energy-efficient ISAC system for V2X networks. Specifically, we first model the dynamics of V2X networks using a Markov Decision Process (MDP) framework. This framework allows the RSU to develop a one-step beamforming scheme based solely on its current sensing data, significantly reducing beamforming overhead compared to existing methods~\cite{liu2020radar,10561505,liu2022learning,10605608,10304580}. Meanwhile, it also eliminates the need for explicit CSI and reduces reliance on pilot signals, thereby enhancing the system's practicality in dynamic and uncertain scenario. To overcome the limitations of conventional tracking-based methods such as EKF and UKF, we introduce a model-free deep reinforcement learning (DRL) algorithm. This algorithm enables the RSU to learn adaptive control policies through interaction with the environment, allowing it to effectively handle nonlinear vehicle dynamics, stochastic noise, and partial observability. To further improve communication performance and sensing accuracy in dynamic V2X network, we jointly optimize the beamforming and power allocation problems. This approach ensures that the optimized policy can adaptively respond to time-varying channel conditions and changing vehicle states, providing considerable communication rates and sensing accuracy in uncertain and dynamic V2X environments.
In addition, we introduce a novel neural network framework, spiking neural networks (SNNs), and integrate it with the proposed DRL algorithm. Leveraging their event-driven nature and asynchronous processing capabilities, SNNs excel in managing spatiotemporal patterns, enhancing real-time decision-making while significantly reducing energy consumption. These features enable SNN-driven DRL algorithm well-suited for ISAC tasks in energy-constrained automotive systems, meeting the safety-critical and latency-sensitive demands of V2X applications. Our contributions are summarized as follows:  
\begin{itemize}
    \item We develop a novel stochastic optimization framework that jointly optimizes beamforming and power allocation schemes to maximize sum communication rates while ensuring sensing accuracy in ISAC-assisted V2X networks. By reformulating this optimization problem as an MDP, we then design a one-step beamforming scheme from the RSU. This scheme executes beamforming based solely on the RSU's current sensing data for surrounding vehicles, significantly reducing beamforming overhead. 

    \item We introduce a highly effective learning-based algorithm that enables the RSU to quickly find the optimal policy by utilizing the capabilities of model-free DRL and the actor-critic framework with a policy-clipping technique. This learning-based approach eliminates the need for explicit model linearization and assumptions about known system noise statistics, as in~\cite{liu2020radar,10659350,10561505}, by learning optimal decision-making strategies directly from environmental interaction data and observed states (i.e., sensing data), enabling the RSU to make real-time and adaptive decisions in the dynamic and uncertain V2X environments.

    \item To alleviate the energy consumption of learning-based algorithms, we propose an advanced DRL algorithm enhanced with energy-efficient SNNs. By replacing traditional neural networks with SNNs, the proposed SNN-driven DRL algorithm significantly reduces the energy consumption of neural network operations. Furthermore, leveraging the unique spatiotemporal dynamics of SNNs, the algorithm achieves enhanced performance in both convergence speed and decision-making.
    
    \item We conduct comprehensive simulations to evaluate the efficiency of the proposed method. The results show that the SNN-driven DRL algorithm achieves energy savings of 5.15 times during model training and 7.125 times during inference compared to conventional approaches. In addition, the proposed scheme exhibits superior performance in terms of enhanced communication rates and improved sensing accuracy.
\end{itemize}

The rest of this paper is organized as follows. Section \ref{Se:system model} introduces the system model and provides a detailed problem formulation. Section \ref{sec3} explains the MDP framework and the one-step beamforming scheme. In Section \ref{sec4}, we propose the intelligent and energy-efficient learning algorithm. Sections \ref{PE} and \ref{conclusion} present comprehensive evaluations and the conclusion of this work, respectively.

% \newpage
% \vspace{-7pt}
\section{System Model and Problem Formulation}\label{Se:system model}
\begin{figure}[t!]
    % \vspace{-5pt}
    \centering
    \includegraphics[width=0.48\textwidth]{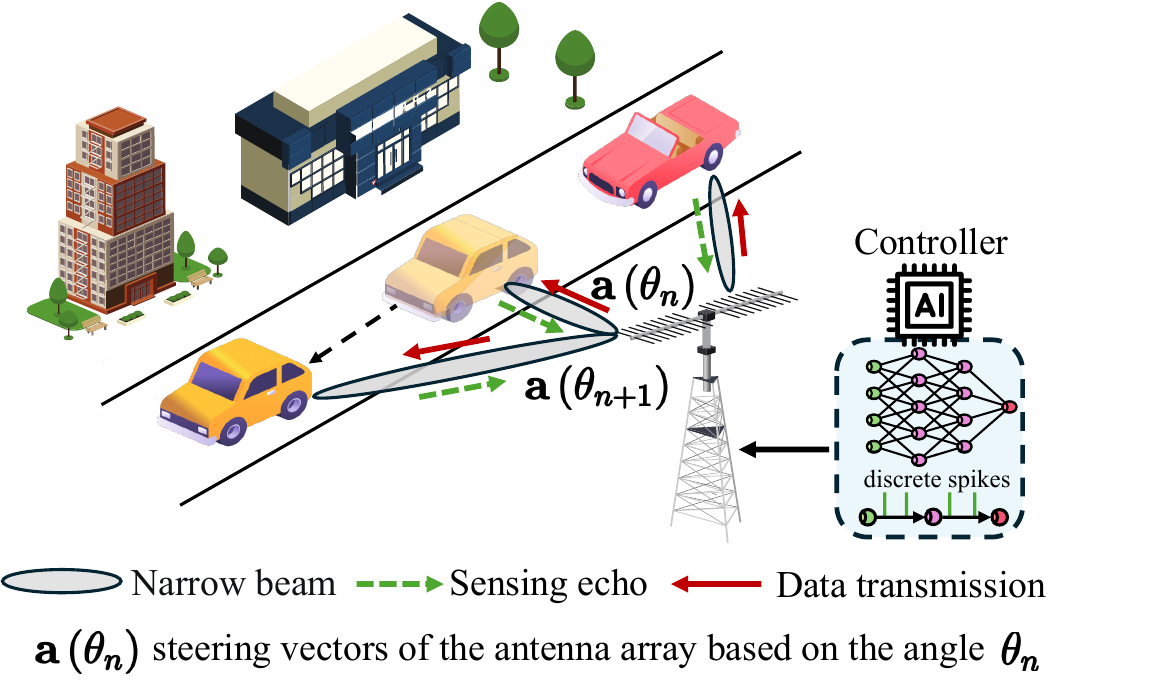}
    \caption{The AI-based ISAC-assisted V2X networks. By leveraging the discrete spikes of SNNs for decision-making, the RSU significantly reduces energy consumption.}
    \label{Fig: scenario}
    % \vspace{-10pt}
\end{figure}

As shown in Fig.~\ref{Fig: scenario}, we consider an ISAC-assisted V2X network consisting of an RSU controlled by AI algorithm and multiple vehicles, where the vehicles are driving along a straight, single-lane road parallel to the antenna array of RSU and each vehicle represented by $\mathcal{K} =\left\{ 1, 2,\ldots , K \right\} $. The RSU is equipped with massive MIMO uniform linear arrays (ULA) antennas, comprising $N_{\text{TA}}$ transmission antennas and $N_{\text{RA}}$ receive antennas, to transmit and receive ISAC signals containing both radar (i.e., sensing) and communication components for the vehicles with single antenna. The operation processes of this system are as follows: (1) The RSU sends the ISAC signal through the narrow beam to these vehicles; (2) The echo signals from sensing signals are reflected back to the RSU by the vehicles; (3) The RSU estimates the vehicles' positions and optimizes the beamforming based on the reflected sensing signals; (4) The ISAC signals are transmitted using the optimized beamforming schemes. Through this iterative process, the RSU provides an efficient beamforming scheme, ensuring that the communication components of the ISAC signals are effectively received by the vehicles, thereby achieving sensing-assisted communication. The specific sensing and communication models are detailed below.
% \vspace{-10pt}
\subsection{Sensing Model}\label{sensing model}
% \vspace{-5pt}
Let $x_{k,n}\left( t \right) $ denote the downlink ISAC signal transmitted to the vehicle-$k$ at time instant $t$ within the $n$-th time slot, where $n\in \left\{ 1,2,\dots ,N \right\} $, and the ISAC signal vector for all $K$ vehicles can be expressed as $\boldsymbol{x}_n\left( t \right) =\left[ x_{1,n}\left( t \right) ,x_{2,n}\left( t \right) ,...,x_{K,n}\left( t \right) \right] ^T\in \mathbb{C} ^{K\times 1}$. Therefore, the transmitted signal through the $N_{\text{TA}}$ transmission antennas at the RSU can be expressed as:
\begin{equation}
\tilde{\boldsymbol{x}}_n\left( t \right) =\mathbf{F}_n\boldsymbol{x}_n\left( t \right) \in \mathbb{C} ^{N_{\text{TA}}\times 1},
\end{equation}
where $\mathbf{F}_n=\left[ \mathbf{f}_{1,n},\mathbf{f}_{2,n},...,\mathbf{f}_{K,n} \right] $ is the transmit beamforming matrix with $\mathbf{f}_{k,n}\in \mathbb{C} ^{N_{\text{TA}}\times 1}$ at $n$-th time slot. After a round-trip, the reflected echo signal received at the RSU is given by:
\begin{equation}
\begin{aligned}
\boldsymbol{y}_n\left( t \right) = ~&\mathcal{E} \sum_{k=1}^K \sqrt{p_{k,n}} \beta_{k,n} e^{j2\pi \mu_{k,n} t} \mathbf{b}\left( \theta_{k,n} \right) \mathbf{a}^H\left( \theta_{k,n} \right) \\
&\tilde{\boldsymbol{x}}_n\left( t-\tau_{k,n} \right) 
+ \mathbf{z}\left( t \right),
\end{aligned}~
\label{sumisac}
\end{equation}
where $\mathcal{E} =\sqrt{N_{\text{TA}}N_{\text{RA}}}$ and $p_{k,n}$ represent the array gain factor for sensing signal and the transmit power of the RSU for beam-$k$ at time slot $n$, respectively. $\beta _{k,n}=\kappa /\left( 2d_{k,n} \right) $ is the reflection coefficient of vehicle-$k$ at time slot $n$, where $\kappa $ represents the fading coefficient and $d_{k,n}$ is the distance between vehicle-$k$ and the RSU at time slot $n$. Besides, $\theta _{k,n}$, $\mu_{k,n}$, and $\tau_{k,n}$ denote the angle between the vehicle-$k$ and the RSU, the doppler frequency, and the round-trip time-delay of echo signal at time slot $n$, respectively. $\mathbf{z}\left( t \right) \in \mathbb{C} ^{N_{\text{RA}}\times 1}$ denotes the complex additive white Gaussian noise with zero mean and variance of $\sigma ^2$, i.e., $\mathbf{z}\sim \mathcal{C} \mathcal{N} \left( 0,\sigma ^2 \right) $. Furthermore, ${\mathbf{a}}\left( {{\theta_{k,n}}} \right)$ and ${\mathbf{b}}\left( {{\theta_{k,n}}} \right)$ are transmit and receive steering vectors of the antenna array of the RSU, respectively, which can be expressed as follows:
\begin{equation}\label{eq4}
\mathbf{a}\left( \theta _{k,n} \right) =\sqrt{\frac{1}{N_{\text{TA}}}}\left[ 1,e^{-j\pi \cos \theta _{k,n}},...,e^{-j\pi \left( N_{\text{TA}}-1 \right) \cos \theta _{k,n}} \right] ^T,
\end{equation}
\begin{equation}\label{eq5}
\mathbf{b}\left( \theta _{k,n} \right) =\sqrt{\frac{1}{N_{\text{RA}}}}\left[ 1,e^{-j\pi \cos \theta _{k,n}},...,e^{-j\pi \left( N_{\text{RA}}-1 \right) \cos \theta _{k,n}} \right] ^T.
% \vspace{-4pt}
\end{equation}

Note that the reflected echo signal in (\ref{sumisac}) consists of multiple sub-echoes originating from different vehicles, leading to unavoidable inter-beam interference. To address this issue, we introduce the following lemma based on the theory of massive MIMO.

\textit{Lemma 1: The steering vectors corresponding to different angles are asymptotically orthogonal, i.e., $\left| \mathbf{b}^H\left( \theta _{k,n} \right) \mathbf{b}\left( \theta _{k',n} \right) \right|\approx 0,~\forall k\ne k'$.}

\textit{Proof: Please refer to}~\cite{marzetta2016fundamentals,liu2020radar}.

Based on \textit{Lemma 1}, 
the RSU can distinguish multiple vehicles based on their angle-of-arrival (AoA), enabling effective spatial separation and suppression of self-interference~\cite{ngo2015massive}. To this end, a spatial filtering operation~\cite{richards2005fundamentals} can be applied, and the received echo signal at the RSU reflected by vehicle-$k$, denoted as $y_{k,n}\left( t \right)$, can be extracted from (\ref{sumisac}) and expressed as:
\begin{equation}
\begin{aligned}
y_{k,n}\left( t \right) =&\mathbf{b}^H( \hat{\theta}_{k,n} ) \boldsymbol{y}_{n}\left( t \right) 
\\
=&\mathcal{E} \sqrt{p_{k,n}}\beta _{k,n}e^{j2\pi \mu _{k,n}t}\mathbf{a}^H\left( \theta _{k,n} \right) \tilde{\boldsymbol{x}}_n\left( t-\tau _{k,n} \right) 
\\
&+z_{k,n}\left( t \right) 
\end{aligned},
\label{eq:echo for vehicle}
\end{equation}
where $\mathbf{b}^H( \hat{\theta}_{k,n}) \boldsymbol{y}_{n}\left( t \right) $ represents the spatial filtering operation based on the estimated angle $\hat{\theta}_{k,n}$ ($\hat{\theta}_{k,n}\approx{\theta}_{k,n}$) of the receive beamforming vector. $z_{k,n}(t) = \mathbf{b}^H( \hat{\theta}_{k,n} ) \mathbf{z}(t)$ with $z_{k,n}\left( t \right) \sim \mathcal{C} \mathcal{N} \left( 0,\sigma _{z}^{2} \right) $ denotes the noise vector.

After extracting the received echo signal for vehicles, the RSU is able to estimate their states, and subsequently update the beamforming scheme based on these estimated states information. Specifically, matched filtering can be applied to maximize the signal-to-noise ratio (SNR), thereby enhancing the detectability of the echo signal~\cite{van2002optimum}. This process can be expressed as:
% \vspace{-5pt}
\begin{equation}
\begin{aligned}
&\left\{\tilde{\tau}_{k,n},\tilde{\mu}_{k,n} \right\} \\
&=\underset{\tau _{k,n},\mu _{k,n}}{\mathrm{arg}\max}\left| \int_0^{\Delta T_m}{y_{k,n}\left( t \right) x_{k,n}^{*}\left( t-\tau \right) e^{-j2\pi \mu t}dt} \right|^2,  
\end{aligned}
\label{matched-filtering}
% \vspace{-5pt}
\end{equation}
where $\tilde{\mu}_{k,n}$ and $\tilde{\tau}_{k,n}$ denote the measured doppler frequency and time-delay after matched-filtering operation, $\Delta T_m$ is the duration of the observation window used for sensing, and $x^*$ is the complex conjugate of $x$. As a result, the RSU is able to estimate the states of vehicles based on these measurement parameters. The mathematical representations of the measured doppler frequency $\tilde{\mu}_{k,n}$, estimated distance $\hat{d}_{k,n}$, time-delay $\tilde{\tau}_{k,n}$, and the velocity $\hat{v}_{k,n}$ of vehicles-$k$ are given by:
\begin{equation}
\tilde{\tau} _{k,n}=\frac{2\hat{d}_{k,n}}{c}+z_{\tau_k,n},~~ \tilde{\mu} _{k,n}=\frac{2\hat{v}_{k,n}\cos \theta _{k,n}f_c}{c}+z_{\mu_k,n}  ,
\label{measure}
% \vspace{-5pt}
\end{equation}
where $z_{\tau_k,n}$ and $z_{\mu_k,n}$ represent the measurement noise with zero mean and variance of $\sigma _{\tau_k,n}^{2}$ and $\sigma _{\mu_k,n}^{2}$ at time slot $n$, respectively, $c$ and $f_c$ are the speed of light and the carrier frequency. Specifically, $\sigma _{\tau_k,n}^{2}$ and $\sigma _{\mu_k,n}^{2}$ are inversely proportional to the received signal-to-interference-plus-noise ratio (SINR) of the received echo signal (\ref{eq:echo for vehicle})~\cite{liu2020radar,richards2005fundamentals}, which can be expressed as:
% \vspace{-5pt}
\begin{equation}
\mathrm{SINR}_{k,n}= \frac{\mathcal{E}^2p_{k,n}|\beta _{k,n}|^2|\mathbf{a}^H(\theta _{k,n})\mathbf{f}_{k,n}|^2}{\sum_{i\ne k}^K{\mathcal{E}^2p_{i,n}}|\beta _{i,n}|^2\left| \mathbf{a}^H(\theta _{k,n})\mathbf{f}_{i,n} \right|^2+\sigma _{z}^{2}}.
\label{eq:sensing sinr}
\end{equation}
As a result, the variance of noise terms $z_{\tau_{k},n}$ and $z_{\mu_{k},n}$ in (\ref{measure}) can be calculated by~\cite{liu2020radar,richards2005fundamentals}:
% \vspace{-5pt}
\begin{equation}
\sigma _{\tau_k,n}^{2}=\frac{\alpha _{\tau}^{2}\left( \sum_{i\ne k}^K\mathcal{E}^2{p_{i,n}}|\beta _{i,n}|^2\left| \mathbf{a}^H(\theta _{k,n})\mathbf{f}_{i,n} \right|^2+\sigma _{z}^{2} \right)}{\mathcal{E}^2p_{k,n}|\beta _{k,n}|^2|\mathbf{a}^H(\theta _{k,n})\mathbf{f}_{k,n}|^2},
\end{equation}
\begin{equation}
 \sigma _{\mu_k,n}^{2}=\frac{\alpha _{\mu}^{2}\left( \sum_{i\ne k}^K{\mathcal{E}^2p_{i,n}}|\beta _{i,n}|^2\left| \mathbf{a}^H(\theta _{k,n})\mathbf{f}_{i,n} \right|^2+\sigma _{z}^{2} \right)}{\mathcal{E}^2p_{k,n}|\beta _{k,n}|^2|\mathbf{a}^H(\theta _{k,n})\mathbf{f}_{k,n}|^2},   
\end{equation}
where $\alpha _{\tau}^{2}$ and $\alpha _{\mu}^{2}$ are the constant related to system configuration and the signal design. 

It is worth noting that the RSU acts as a pure mono-static radar when there are no vehicles in its coverage area. Once a vehicle enters its coverage area, the RSU can estimate the vehicle’s states and quickly perform beam alignment for data transmission. The communication model is then presented in the following. 

\vspace{-5pt}
\subsection{Communication Model}
\vspace{-5pt}
The received communication signal for vehicle-$k$ at time slot $n$, denoted by $\mathcal{C} _{k,n}$, can be expressed as:
% \vspace{-5pt}
\begin{equation}
\begin{aligned}
\mathcal{C} _{k,n}\left( t \right) =&\bar{\mathcal{E}}\sqrt{p_{k,n}}\alpha _{k,n}e^{j2\pi \mu _{k,n}^{\prime}t}\mathbf{a}^H\left( \theta _{k,n} \right) \sum_{k=1}^K{\mathbf{f}_{k,n}x_{k,n}\left( t \right)}\\
&+z_c\left( t \right)
\end{aligned}
\label{eq:communication for vehicle}~,
\end{equation}
where $\bar{\mathcal{E}} =\sqrt{N_{\text{TA}}}$ denotes the array gain factor for communication signal, $\alpha _{k,n}=\sqrt{\alpha _0\left( d_{k,n}/d_o \right) ^{-\varrho}}$ represents the path loss coefficient, where $\alpha _0$ and $\varrho $ denote the path loss at reference distance $d_0$ and the path loss exponent, respectively. Additionally, $z_c\left( t \right) \sim \mathcal{C} \mathcal{N} \left( 0,\sigma _{c}^{2} \right) $ denotes the noise term. Note that the path loss coefficient $\alpha _{k,n}$ and doppler frequency $\mu _{k,n}^{\prime}=v_{k,n}\cos \theta _{k,n}f_c/c$ in (\ref{eq:communication for vehicle}) differs from $\beta _{k,n}$ in (\ref{sumisac}) and $\tilde{\mu} _{k,n}$ in (\ref{measure}), respectively, because the communication signal only involves a one-way transmission, whereas the sensing signal undergoes a round-trip propagation.

Therefore, given the beamforming scheme $\mathbf{f}_{k,n}$ and the allocated transmit power $p_{k,n}$, the communication signal's SINR for vehicle-$k$, denoted by $\gamma _{k,n}$, is given by:
\begin{equation}
\gamma _{k,n}\left( \mathbf{f}_{k,n},p_{k,n} \right) =\frac{\bar{\mathcal{E}}^2p_{k,n}\left|\alpha _{k,n}\mathbf{a}^H\left( \theta _{k,n} \right) \mathbf{f}_{k,n} \right|^2}{\sum_{i\ne k}^K{\bar{\mathcal{E}}^2p_{i,n}\left| \alpha _{k,n}\mathbf{a}^H\left( \theta _{k,n} \right) \mathbf{f}_{i,n} \right|^2}+\sigma _{c}^{2}},
\label{SINR of communication}
\end{equation}
and then the achievable transmit sum-rate of all the vehicles in the ISAC-assisted vehicle network system can be expressed as:
% \vspace{-5pt}
\begin{equation}
R=\sum_{k=1}^K{R_k}=\sum_{k=1}^K{\log _2\left( 1+\gamma _{k,n}\left( \mathbf{f}_{k,n},p_{k,n} \right) \right)}.
\label{sum-rate}
% \vspace{-4pt}
\end{equation}

It can be observed that the achievable sum-rate is highly dependent on the beamforming and power allocation schemes. Moreover, the efficacy of beamforming can be enhanced by accurate sensing, thereby enabling sensing-assisted communication. However, the RSU is unable to directly evaluate its sensing performance, i.e., the mean square error (RMSE), because the true state information of the vehicles is unknown. To address this challenge, we introduce the Cramér-Rao Lower Bound (CRLB)~\cite{9652071} in the following subsection.

% \vspace{-12pt}
\subsection{Cramér-Rao Lower Bound and Problem Formulation}\label{Problem Formulation}
Let $\mathbf{x}_{k,n}=\left[ \theta _{k,n},d_{k,n},v_{k,n} \right] ^T$ represent the states of the vehicle-$k$ at time slot $n$, its state evolution model can be expressed as~\cite{liu2020radar}:
\begin{equation}
\begin{cases}
	\theta _{k,n}=\theta _{k,n-1}+d_{k,n-1}^{-1}v_{n-1}\Delta T\sin \theta _{k,n-1}+\omega _{\theta},\\
	d_{k,n}=d_{k,n-1}-v_{k,n-1}\Delta T\cos \theta _{k,n-1}+\omega _d,\\
	v_{k,n}=v_{k,n-1}+\omega _v,\\
\end{cases}
\end{equation}
where $\Delta T$ is the time duration of each time slot, which differs from the sensing observation window $\Delta T_m$ defined in (\ref{matched-filtering}) ($\Delta T_m \ll \Delta T$)~\cite{9705498,9737357}, $\omega _\theta$, $\omega _d$, and $\omega _v$ represent the noise with zero mean and variances $\sigma _{\theta}^{2}$, $\sigma _{d}^{2}$, and $\sigma _{v}^{2}$, respectively. Note that the RSU is unable to directly collect the vehicles’ states, i.e., the state evolution model of vihicles is unknown to the RSU, therefore, it must measure and estimate these parameters using the sensing signal, as illustrated in Section~\ref{sensing model}.

Let $\mathbf{y}_{k,n}=\left[ \tilde{y}_{k,n},\tilde{\tau}_{k,n}, \tilde{\mu}_{k,n} \right] ^T$ denote the measurement model of the RSU, which maps the vehicle states $\mathbf{x}_{k,n}$ to the corresponding measurements. Specifically, $\tilde{\tau}_{k,n}$ and $\tilde{\mu}_{k,n}$ are the measurement parameters derived from the signal $\tilde{y}_{k,n}$ that experienced a matched filtering operation in (\ref{matched-filtering}). To this end, we first rewrite the right-hand term of (\ref{matched-filtering}) as:
\begin{equation}
\begin{aligned}
	&\tilde{y}_{k,n}=\int_0^{\varDelta T_m}{y}_{k,n}\left( t \right)x_{k,n}^{*}\left( t-\tilde{\tau}_{k,n} \right) e^{-j2\pi \tilde{\mu}_{k,n}t}dt\\
	=&\mathcal{E} \sqrt{p_{k,n}}\beta _{k,n} \mathbf{a}^H\left( \theta _{k,n} \right) \mathbf{f}_{k,n}\times\\
	&\int_0^{\varDelta T_m}{x_{k,n}\left( t-\tilde{\tau}_{k,n} \right) x_{k,n}^{*}\left( t-\tilde{\tau}_{k,n} \right) e^{j2\pi \left( \mu _{k,n}t-\tilde{\mu}_{k,n}t \right)}dt}\\
	&+\int_0^{\varDelta T_m}{z_{k,n}\left( t \right) x_{k,n}^{*}\left( t-\tilde{\tau}_{k,n} \right) e^{-j2\pi \tilde{\mu}_{k,n}t}dt}\\
	=&\mathcal{E}\sqrt{p_{k,n}} \beta _{k,n} \xi\mathbf{a}^H\left( \theta _{k,n} \right) \mathbf{f}_{k,n}  +\tilde{z}_{k,n}\\
\end{aligned},
\label{eq:echo for recovery}
\end{equation}
where $\xi$ represents the matched-filtering gain, and $\tilde{z}_{k,n}\sim \mathcal{C} \mathcal{N} \left( 0,\sigma _{y_k}^{2} \right) $ is the noise with zero mean and $\sigma _{y_k}^{2}$ being the variance. Then, the mapping function between $\mathbf{y}_{k,n}$ and $\mathbf{x}_{k,n}$ is given by:
\begin{equation}
\mathbf{y}_{k,n}=\mathbf{g}\left( \mathbf{x}_{k,n} \right) +\mathbf{u}_{k,n},
\label{measuremodel}
\end{equation}
where $\mathbf{g}\left( \cdot \right) $ is defined by  (\ref{measure}) and (\ref{eq:echo for recovery}), $\mathbf{u}_{k,n}=\left[ \tilde{z}_{k,n},z_{\tau_k,n},z_{\mu_k,n} \right] ^T$, and $\mathbf{y}_{k,n}\sim \mathcal{C} \mathcal{N} \left( \mathbf{g}\left( \mathbf{x}_{k,n} \right) ,\mathbf{Q} \right) $, where $\mathbf{Q}=\mathrm{diag}\left( \left[ \sigma _{y_k}^{2},\sigma _{\tau _k,n}^{2},\sigma _{\mu _k,n}^{2} \right] \right) $ is the covariance matrix. As a result, the RSU is able to estimate the states of vehicles $\hat{\mathbf{x}}_{k,n}=\left[ \hat{\theta}_{k,n},\hat{d}_{k,n},\hat{v}_{k,n} \right] ^T$ from the sensing signal.

As aforementioned, we introduce the CRLB to evaluate the measurement accuracy due to the unknown real states of vehicles. The CRLB provides the lower bound for the variance of unbiased estimators in target estimation performance~\cite{kay1993fundamentals}, i.e., $\mathbf{x}_{k,n}$ in this work. Specifically, the CRLB is defined as the inverse of the \textit{Fisher Information Matrix} $\mathbf{FIM}\left( \mathbf{x}_{k,n} \right)$, which can be expressed as~\cite{kay1993fundamentals}:
\begin{equation}
\begin{aligned}
\mathbf{FIM}\left( \mathbf{x}_{k,n} \right) &=\mathbb{E} \left[ \frac{\partial ^2\ln  p\left( \mathbf{y}_{k,n}\left| \mathbf{x}_{k,n} \right. \right)}{\partial ^2\mathbf{x}_{k,n}} \right] \\
&=\mathbb{E} \left[ \left( \frac{\partial \mathbf{g}\left( \mathbf{x}_{k,n} \right)}{\partial \mathbf{x}_{k,n}} \right) ^H\mathbf{Q}^{-1}\left( \frac{\partial \mathbf{g}\left( \mathbf{x}_{k,n} \right)}{\partial \mathbf{x}_{k,n}} \right) \right] ,
\end{aligned}
\end{equation}
where $p\left( \mathbf{y}_{k,n}\left| \mathbf{x}_{k,n} \right. \right) $ is the conditional probability density function (PDF) of $\mathbf{y}_{k,n}$ given $\mathbf{x}_{k,n}$~\cite{9652071,kay1993fundamentals}.
Therefore, the lower bound on the variance of an unbiased estimator, i.e., the CRLB, can be expressed as:
\begin{equation}
\mathbb{E} \left[ \left( \hat{\mathbf{x}}_{k,n}-\mathbf{x}_{k,n} \right) \left( \hat{\mathbf{x}}_{k,n}-\mathbf{x}_{k,n} \right) ^H \right] \succeq \mathbf{FIM}^{-1}\left( \mathbf{x}_{k,n} \right).
\end{equation}
Note that the variance is equal to the RMSE due to we only consider the unbiased estimator.
% \footnotemark\footnotetext{For the RSU, it first estimates the radial velocity by measuring the Doppler frequency shift and then calculate the vehicle's velocity using the estimated radial velocity and the angle $\theta$. Thus we can assume the Doppler frequency shift is only determined by the radial velocity, i.e., $\frac{\partial \mu}{\partial \theta} = 0$.}
Besides, we have:
\begin{equation}
\frac{\partial \mathbf{g}\left( \mathbf{x}_{k,n} \right)}{\partial \mathbf{x}_{k,n}}=\left[ \begin{matrix}
	\frac{\partial \tilde{y}_{k,n}}{\partial \theta _{k,n}}&		0&		0\\
	0&		\frac{2}{c}&		0\\
	0&		0&		\frac{2f_c\cos \theta}{c}\\
\end{matrix} \right] \in \mathbb{C} ^{\left( N_{\text{RA}}+2 \right) \times \left( N_{\text{RA}}+2 \right)}.
\label{g/x}
\end{equation}
According to the CRLB theorem~\cite{9652071,kay1993fundamentals}, given the estimated angle $\hat{\theta}_{k,n}$ and distance $\hat{d}_{k,n}$, their RMSEs are respectively bounded by:
\begin{equation}
\mathbb{E} \left[ \left( \hat{\theta}_{k,n}-\theta _{k,n} \right) ^2 \right] \geqslant \mathrm{CRLB}_{\theta}\left( \theta _{k,n},\mathbf{f}_{k,n},p_{k,n} \right)\triangleq \mathbf{FIM}_{11}^{-1} ,
\end{equation}
\begin{equation}
\mathbb{E} \left[ \left( \hat{d}_{k,n}-d_{k,n} \right) ^2 \right] \geqslant \mathrm{CRLB}_d\left( d_{k,n},\mathbf{f}_{k,n},p_{k,n} \right)\triangleq \mathbf{FIM}_{22}^{-1} ,
\end{equation}
where $\mathbf{FIM}^{-1}_{ij}$ denotes the $i$-th row and the $j$-th column element of $\mathbf{FIM}^{-1}$. Furthermore, given the beamforming matrix $\mathbf{f}_{k,n}$ and the allocated transmit power $p_{k,n}$, the CRLB of $\theta_{k,n}$ and $d_{k,n}$ can be respectively expressed as follows:
\begin{equation}
\mathrm{CRLB}_{\theta}\left( \theta _{k,n},\mathbf{f}_{k,n},p_{k,n} \right) =\left[ \frac{1}{\sigma _{y_k}^{2}}\left( \frac{\partial \tilde{y}_{k,n}}{\partial \theta _{k,n}} \right) ^H\frac{\partial \tilde{y}_{k,n}}{\partial \theta _{k,n}} \right] ^{-1},
\label{eq:crlb-theta}
\end{equation}
and
\begin{equation}
\mathrm{CRLB}_d\left( d_{k,n},\mathbf{f}_{k,n},p_{k,n} \right) =\left[ \frac{1}{\sigma _{\tau_k}^{2}}\left( \frac{2}{c} \right) ^2 \right] ^{-1}.
\end{equation}
Through introducing the CRLB, we are able to quantify the lower bound of the variance that any unbiased estimator might reach when estimating $\theta_{k,n}$ and $d_{k,n}$. 
In other words, it allows us to directly optimize the beamforming and power allocation schemes without relying on the estimation accuracy (i.e., the RMSE). 

As illustrated above, this work aims to leverage the sensing signal to reduce the complexity of beamforming training, thereby assisting in optimizing the beamforming design, and ultimately enhancing communication performance. The RSU's objective is to maximize the achievable sum-rate across all vehicles while ensuring sensing performance through optimal beamforming design and power allocation. Therefore, the optimization problem can be formulated as follows:
\begin{subequations}\label{eqn:optimization-problem}
\begin{align}
  \textbf{P1:}~ &\underset{\{\mathbf{F}_n,~\boldsymbol{p}_n\}_{n=1}^{N}}{\max}\,\mathbb{E} \left[ \frac{1}{N}\sum_{n=1}^N{\sum_{k=1}^K{\log _2\left( 1+\gamma _{k,n}\left( \mathbf{f}_{k,n},p_{k,n} \right) \right)}} \right] , \tag{\ref{eqn:optimization-problem}} \\
\text{s.t.}\quad & \mathbb{E} \left[ \frac{1}{N}\frac{1}{K}\sum_{k=1}^K{\mathrm{CRLB}_{\theta}\left( \theta _{k,n},\mathbf{f}_{k,n},p_{k,n} \right)} \right] \leqslant \epsilon _{\theta}, \label{eqn:pbm-const-1} \\
& \mathbb{E} \left[ \frac{1}{N}\frac{1}{K}\sum_{k=1}^K{\mathrm{CRLB}_{d}\left( d_{k,n},\mathbf{f}_{k,n},p_{k,n} \right)} \right] \leqslant \epsilon _d, \label{eqn:pbm-const-2} \\
& \sum_{k=1}^K{p_{k,n}}\leqslant P_{\max} ~~\text{and} ~~p_{k,n}>0 ~~~\forall k, \label{eqn:pbm-const-3}
\end{align}
\end{subequations}
where $\boldsymbol{p}_n=\left[ p_{1,n},p_{2,n},\dots ,p_{K,n} \right] ^T\in \mathbb{C} ^{K\times 1}$ is the allocated power vector, $\epsilon _\theta$ and $\epsilon _d$ are the maximum tolerable CRLB thresholds for reliable sensing, and $P_{\max}$ represents the RSU's maximum transmit power. Specifically, the optimization problem is designed to enable the RSU to maximize the expected long-term achievable sum-rate, while simultaneously ensuring that the expected sensing accuracy $\mathrm{CRLB}_{\theta}$ and $\mathrm{CRLB}_{d}$ remains within acceptable thresholds across all vehicles and time slots.

Optimizing \textbf{P1} presents significant challenges due to the highly dynamic and complex nature of the V2X environment, characterized by time-varying and unknown channels between vehicles and RSUs, as well as real-time variations in vehicle positions. These factors make the sum-rate dependent on multiple variables, such as estimation accuracy, channel conditions, power allocation, and interference management. Moreover, the problem is inherently non-convex, further complicating the derivation of optimal solutions. In this situation, classical filtering methods such as EKF and UKF rely on ideal assumptions (e.g., known noise statistics, full observability) and are designed for state estimation rather than joint optimization. They are unable to coordinate beamforming and power allocation decisions across multiple vehicles or adapt to long-term dynamics and partial observability. Similarly, codebook-based beamforming schemes are typically static and optimized for single-user communication scenarios. As such, they are limited in their ability to accommodate joint communication and sensing constraints, lack support for coordinated beamforming across multiple vehicles, and are generally ineffective in highly dynamic environments due to their limited adaptability.

To effectively tackle these challenges, we reformulate the dynamic environment as an MDP and introduce an advanced model-free DRL algorithm in the next section. This innovative approach empowers the system to make efficient, data-driven decisions and adapt dynamically in real-time, ensuring optimal performance even in complex and rapidly changing environments.

\section{Proposed Beamforming Scheme and Problem Transformation based on MDP framework}\label{sec3}
% In this Section, we first introduce the MDP framework to reformulate the problems, we then develop a beamforming scheme based on the MDP framework.
As discussed in Section \ref{Problem Formulation}, the operating environment for ISAC-assisted V2X networks is inherently unpredictable and dynamic, posing significant challenges for conventional optimization algorithms. DRL algorithms have emerged as a powerful solution~\cite{10742922,ppo,arulkumaran2017deep}. By continuously interacting with and adapting to these complex environments, DRL algorithms effectively manage uncertainty and enhance decision-making through their integration of randomness and exploration strategies. Moreover, the proficiency of DRL in capturing and representing complex nonlinear relationships makes it indispensable for these settings. Specifically, model-free DRL algorithms excel by learning optimal strategies directly via interacting with the surrounding environment, avoiding the need for a predefined model. This capacity allows them to adaptively refine their policies in real-time, providing a significant advantage in managing the intricacies of V2X networks where dynamics are significant complex to model explicitly. Therefore, we develop a DRL algorithm to address the challenges of beamforming design and power allocation presented in the ISAC-assisted vehicle network environment. To this end, we first model the dynamics of the vehicle network using an MDP framework in Section~\ref{mdp}. We then illustrate the proposed beamforming scheme based on the MDP framework in Section~\ref{Proposed Beamforming Scheme}. Furthermore, we reformulate the \textbf{P1} into a stochastic optimization problem that can be solved by the DRL in Section~\ref{Problem Transferring}.
\subsection{MDP Framework for the ISAC-assisted V2X Network}\label{mdp}
As discussed above, directly optimizing the considered ISAC-assisted V2X network is challenging due to its high dynamics and uncertainty. Therefore, we first transform it into an MDP framework, which provides a mathematical foundation for sequential decision-making under uncertainty. This framework allows the RSU to adaptively optimize its beamforming and power allocation schemes as the network environment evolves.

Let a tuple $\left( \mathcal{S} , \mathcal{A} , \mathcal{P} , r, G \right) $ represent the MDP, where $\mathcal{S}$ and $\mathcal{A}$ are the state and action spaces, respectively, $\mathcal{P}$ denotes the state transition probability distribution, $r$ and $G \in \left( 0,1 \right) $ are the reward function and the discount factor of long-term reward, respectively. The
corresponding elements in MDP are detailed as follows:
\subsubsection{State Space}
At each time slot $n$, the state space of the ISAC-assisted vehicle network is defined by:
\begin{equation}
\mathcal{S} =\left\{ \left\{ \hat{\theta}_{k,n},\hat{d}_{k,n},\hat{v}_{k,n},\hat{\gamma}_{k,n} \right\} ;1\leqslant k\leqslant K \right\}.
\end{equation}
Note that all variables in the state space are subject to estimation errors due to the gap between sensing performance and the actual parameters. In other words, sensing performance plays a crucial role in accurately modeling the real state space and significantly affects the MDP framework and subsequent optimization processes. For instance, the agent (i.e., the RSU) may not be able to take appropriate actions if there is a significant discrepancy between its estimated state space and the actual state information. This mismatch significantly hinders the RSU's ability to retrieve accurate state information and make optimal decisions, leading to potential degradation in overall system performance. Additionally, the continuous nature of the state space further exacerbates this issue. Fortunately, our proposed advanced DRL algorithm does not rely on perfect estimation performance and is capable of tolerating estimation errors. It enables the RSU to obtain optimal solutions and maintain considerable communication rates even under imperfect estimations, which is detailed in Section~\ref{Model-free DRL}.

\subsubsection{Action Space and Reward Function}
At each time slot $n$, the RSU designs the beamforming scheme and allocates transmission power to each vehicle based on its current observed state space. The action space of the RSU can be defined as follows:
\begin{equation}
\mathcal{A} =\left\{ \mathbf{F}_n,\boldsymbol{p}_n \right\}.
\end{equation}

After taking action $a_n$ from the action space $\mathcal{A}$ at time slot $n$, the RSU receives an immediate reward $r_n(a_n, s_n)$. Based on this reward, the RSU can determine and optimize its subsequent actions in the following steps. Therefore, the objective of the RSU is to maximize its long-term reward. As illustrated in (\ref{eqn:optimization-problem}), this work aims to maximize the communication rates while ensuring sensing performance. Consequently, the reward function, which guides the RSU to improve communication and sensing performance over all time steps (i.e., long-term reward), is defined as:
\begin{equation}
r_n\left( a_n,s_n \right) =\mathbf{1}(R\mathcal{J} -\mathrm{CRLB}_{\theta}-\mathrm{CRLB}_d),
\label{eq:reward function}
\end{equation}
where $\mathbf{1}\left( \cdot \right) $ is an indicator function that equals 1 when (\ref{eqn:pbm-const-1})-(\ref{eqn:pbm-const-3}) are satisfied, and 0 otherwise. $R$ represents the achievable sum-rate defined in (\ref{sum-rate}), and $\mathcal{J} $ is the Jain's fairness index~\cite{sediq2013optimal}, which evaluates whether the achievable rates for all vehicles are distributed equitably and is expressed as
\begin{equation}
\mathcal{J} \left( R_1,R_2,\dots ,R_K \right) =\frac{2\left( \sum_{k=1}^K{R_k} \right) ^2}{K\times \sum_{k=1}^K{R_{k}^{2}}}.
\end{equation}

This reward function encourages the RSU to optimize beamforming and power allocation schemes efficiently, maximizing achievable rates while ensuring fairness in each vehicle’s communication rate. Besides, it enables the exploration of optimal sensing performance, achieving a balance between communication and sensing objectives.

\subsection{Proposed Beamforming Scheme}\label{Proposed Beamforming Scheme}
Based on the MDP framework, we can design an efficient beamforming scheme. To this end, we first introduce the state-of-the-art beamforming schemes. We then highlight the key differences and innovations of our proposed approach.

\subsubsection{Traditional beam training} This scheme involves transmitting downlink pilot signals in multiple directions, measuring signal quality at the receiver, and using uplink feedback to select the beamforming direction. Obviously, frequent updates of downlink pilot signals and uplink feedback introduce significant overhead and degrade communication quality~\cite{8241348}.

% ~\cite{8809900,7400949,6847111}
\subsubsection{Two-stage beam prediction protocol} This protocol integrates radar sensing with communication (i.e., sensing-assisted beamforming), enabling direct adaptation to environmental data without the need for traditional beam training. However, while it eliminates traditional beam training, this approach still requires a disjoint two-stage process, i.e., sensing and CSI prediction, followed by beamforming, leading to significant computational overhead~\cite{liu2020radar,10304580,9246715}.
\subsubsection{Historical channels-based beamforming protocol} This protocol leverages a learning-based method to design beams, relying heavily on historical channel data~\cite{liu2022learning}. Similarly, the continuous update of historical channel data poses significant challenges for the RSU.
\begin{figure}[t!]
    \centering
    \includegraphics[width=0.48\textwidth]{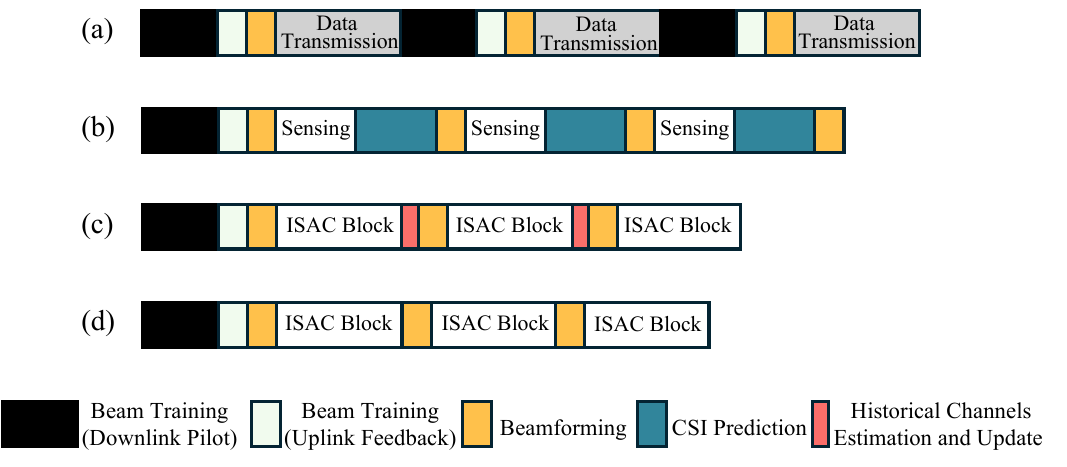}
    \caption{The comparison of different beamforming methods: (a) Traditional beam training, (b) two-stage beam prediction protocol, (c) historical channels-based beamforming protocol, and (d) proposed beamforming scheme. ISAC block includes both data transmission and sensing operations.}
    \label{beamforming schemes}
    % \vspace{-10pt}
\end{figure}

We compare these schemes with our proposed beamforming scheme in Fig.~\ref{beamforming schemes}. Specifically, the ISAC block integrates both communication (i.e., data transmission) and sensing operations (i.e., state information estimation). Unlike these state-of-the-art beamforming schemes above, our proposed beamforming policy is efficient for the following reasons. First, we leverage the sensing signal to assist beamforming, which drastically decreases overhead compared to \textit{Traditional beam training}. Additionally, our scheme operates in a single stage, eliminating the need for multiple stages to first acquire state information, deduce explicit CSI, and then perform beamforming, as required in the \textit{Two-stage beam prediction protocol}. Furthermore, compared to the \textit{Historical channels-based beamforming protocol}, our MDP-based framework eliminates the need for continuous updates of CSI, reducing complexity and resource consumption. As a result, as shown in Fig.~\ref{beamforming schemes}, our proposed beamforming scheme is faster, more resource-efficient, and better suited for dynamic V2X environments than existing methods.

\subsection{Problem Reformulation}\label{Problem Transferring}
Leveraging the MDP framework, we reformulate the original optimization problem in \textbf{P1} into a stochastic optimization problem as follow. 

Let $\pi$ denote a stochastic policy of the RSU (i.e., $\pi :\mathcal{S} \times \mathcal{A} \rightarrow \left[ 0,1 \right] $), which is the probability that taking action $a_n$ given the state $s_n$, i.e., $\pi =\mathbb{P} \left\{ a_n\left| s_n \right. \right\} $. Given the discount of long-term reward $G$, the expected discounted reward of the RSU follows policy $\pi $ is given by:
\begin{equation}
J\left( \pi \right) =\mathbb{E} _{a_n\sim \pi ,s_n\sim \mathcal{P}}\left[ \sum_{n=0}^{\infty}{G^nr_n\left( s_n,a_n \right)} \right],
\end{equation}
where $\mathcal{P} \left( s_{n+1}\left| s_n,a_n \right. \right) $ is the state transition probability distribution that models the dynamics of the environment and is unknown to the RSU. Therefore, the optimization problem in \textbf{P1} can be transformed into finding the optimal policy \(\pi ^*\) that maximizes \(J(\pi )\), expressed as:
\begin{subequations}\label{eqn:maximization-problem}
\begin{align}
&\textbf{P2:}~~~~\mathrm{arg}\underset{\pi}{\max}\,J(\pi )\tag{\ref{eqn:maximization-problem}} \\ 
&\text{s.t.}~ a_n \sim \pi (a_n | s_n), s_{n+1} \sim \mathcal{P}(s_{n+1} | s_n, a_n).  \label{eqn:policy-and-transition}
\end{align}
\label{eq10}
% \vspace{-10pt}
\end{subequations}

By transforming \textbf{P1} into \textbf{P2}, we can apply DRL algorithms to solve the original optimization problem. However, deploying conventional DRL algorithms, such as DQN~\cite{mnih2015human} and DDQN~\cite{wang2016dueling}, presents significant challenges in our scenario. First, traditional DRL algorithms are designed for discrete action spaces and struggle with continuous, dynamic environments like V2X networks. On the other hand, these algorithms require substantial computational resources and energy due to the complexity and frequency of neural network operations. This high computational demand increases energy consumption, raises operational costs, and slows system response times, thereby limiting scalability and practicality in real-world deployment scenarios~\cite{enami2010neural}. 

Consequently, there is an urgent need to design an advanced and energy-efficient DRL algorithm to address these challenges. To this end, we propose an enhanced actor-critic DRL algorithm driven by energy-efficient SNNs in the following section. This innovative approach not only excels at solving complex tasks efficiently but also significantly reduces energy consumption, achieving the dual objectives of optimizing algorithm performance and enhancing energy efficiency.

\section{Proposed Learning Algorithm}\label{sec4}
In this section, we first introduce a model-free and actor-critic DRL algorithm tailored for the dynamic and uncertain environments of V2X networks. We then integrate energy-efficient SNNs with DRL to develop a novel and advanced DRL algorithm to optimize system performance.

\subsection{Model-free DRL Based on Actor-Critic framework}\label{Model-free DRL}
As aforementioned, conventional DRL algorithms struggle to derive the efficient policy $\pi^*$ in the dynamic V2X environment. To address these challenges, we develop an advanced model-free, actor-critic DRL algorithm. Specifically, model-free DRL is a dynamic programming technique designed to solve decision-making problems by learning an optimized policy in dynamic environments~\cite{hoang2023deep}. This approach is particularly well-suited for ISAC-assisted V2I networks as it enhances adaptability to dynamic environments, enables the RSU to make efficient real-time decisions, and effectively handles high-dimensional data. Additionally, by employing two neural networks (i.e., the actor network and critic network), the actor-critic framework allows DRL algorithm to balance bias and variance of training parameters, improve sample efficiency, and ensure stable convergence~\cite{hoang2023deep}.

As shown in the top of Fig.~\ref{Fig:overall}, our proposed learning algorithm employs two specialized neural networks to optimize the RSU's operations, i.e., an actor network and a critic network. The actor network utilizes a parameter vector, denoted as $\boldsymbol{\psi }$ (comprising weights and biases), to generate the policy that guides the RSU's beamforming and power allocation strategies. Concurrently, the critic network, denoted as $\mathbf{\Omega}$, evaluates the performance of the implemented policy by estimating the state-value function. This feedback guides the optimization of $\boldsymbol{\psi}$, enhancing decision-making accuracy and efficiency. Therefore, the optimal policy in (\ref{eq10}) can be approximated as $\pi ^*\gets \pi _{\boldsymbol{\psi }}$ with $\pi _{\boldsymbol{\psi }}\left( a_n\left| s_n \right. \right) =\mathbb{P} \left\{ a_n\left| s_n; \right. \boldsymbol{\psi } \right\} $. Since we focus on the behavior-decision for each action, the time slot index $n$ can be omitted in the following. Note that different from other actor-critc based DRL algorithms~\cite{10742922,10486201,10038571,schulman2017proximal}, the two networks are driven by SNNs rather than conventional neural networks, which are detailed in the Section~\ref{algorithm}. After obtaining the outputs of the two networks, an advantage function $\hat{A}\left( \cdot \right) $ is introduced to evaluate the current policy and guide the policy update of the neural network. Specifically, the advantage function measures whether the action taken is better than the policy's default behavior, which is defined as~\cite{schulman2015high}:
\begin{equation}
\hat{A}\left( s ,a ;\boldsymbol{\psi } \right) =Q \left( s ,a ;\boldsymbol{\psi } \right) -V \left( s ;\mathbf{\Omega } \right) ,
\label{advanf}
\end{equation}
where $Q \left( s ,a ;\boldsymbol{\psi } \right) \,\,$=$\,\,\mathbb{E} _{a\sim \pi _{\boldsymbol{\psi }},s\sim \mathcal{P}}\left[ \sum_{l=0}^{\infty}{G ^lr \left( s_{l}\\,a_{l} \right)} \right] $ is the action-value function,  $V \left( s ;\mathbf{\Omega } \right) =\mathbb{E} _{s \sim \mathcal{P}}\left[ \sum_{l=0}^{\infty}{G^lr \left( s_{l},a_{l} \right)} \right] $ denotes the state-value function, and $l$ is the constant related to time step. Once the value of the advantage function is obtained, the training objective function of actor network can be expressed as~\cite{schulman2017proximal}:
\begin{equation}
J\left( \boldsymbol{\psi } \right)  =\min \left( \frac{\pi _{\boldsymbol{\psi }}\left( a  \left| s   \right. \right)}{\pi _{\boldsymbol{\psi }_{old}}\left( a  \left| s   \right. \right)}\hat{A}  ,\phi \left( \epsilon ,\hat{A}   \right) \right) ,
\label{eq:loss of actor}
\end{equation}
where $\pi _{\boldsymbol{\psi }_{old}}$ denotes the vector of policy parameters before the update. In particular, $\phi(\epsilon, \hat{A})$ represents the policy-clipping technique, which constrains the policy update ratio within a predefined range, ensuring stable gradient updates. By preventing excessively large policy changes, this method stabilizes the training process and enhances the algorithm's performance, which is defined as:
\begin{equation}
\phi \left( \epsilon ,\hat{A}   \right) =\left\{ \begin{array}{c}
	\left( 1+\epsilon \right) \hat{A}  ,\mathrm{if}\hat{A}  \geqslant 0,\\
	\left( 1-\epsilon \right) \hat{A}  ,\mathrm{if}\hat{A}  <0.\\
\end{array} \right. 
\label{eq:policy-clipping}
\end{equation}

% \vspace{-15pt}
\subsection{Enhanced-DRL Through Spiking Neural Network}\label{algorithm}

As aforementioned, the complexity and frequent computations of conventional neural networks impose significant energy demands on the RSU. To address these challenges, we propose an enhanced DRL algorithm driven by SNNs. Unlike traditional neural networks, which propagate information using continuous activation values and typically rely on multiply-and-accumulate (MAC) operations~\cite{masadeh2019input}, SNNs transmit information using discrete spikes of electrical activity. In other words, computation in SNNs is event-driven, meaning neurons perform computations only when spikes occur (represented as binary events, i.e., 0 or 1). This event-driven nature allows SNNs to perform training and inference operations (e.g., forward and back-propagation) utilizing simple accumulation (AC) instead of MAC operations, thus significantly reducing computational complexity and energy consumption~\cite{10742922,fang2021incorporating,9997098}.
As a result, the SNN-driven DRL framework enables RSUs to make decisions with reduced energy consumption due to the discrete, spike-based processing. On the other hand, the temporal nature of spikes allows the RSU to efficiently capture temporal dependencies, thereby improving the training and inference efficiency of the algorithm. The details of the SNN architecture are as follows.
\subsubsection{The framework of SNNs}
\begin{figure}[t]
    \centering
    \includegraphics[width=0.48\textwidth]{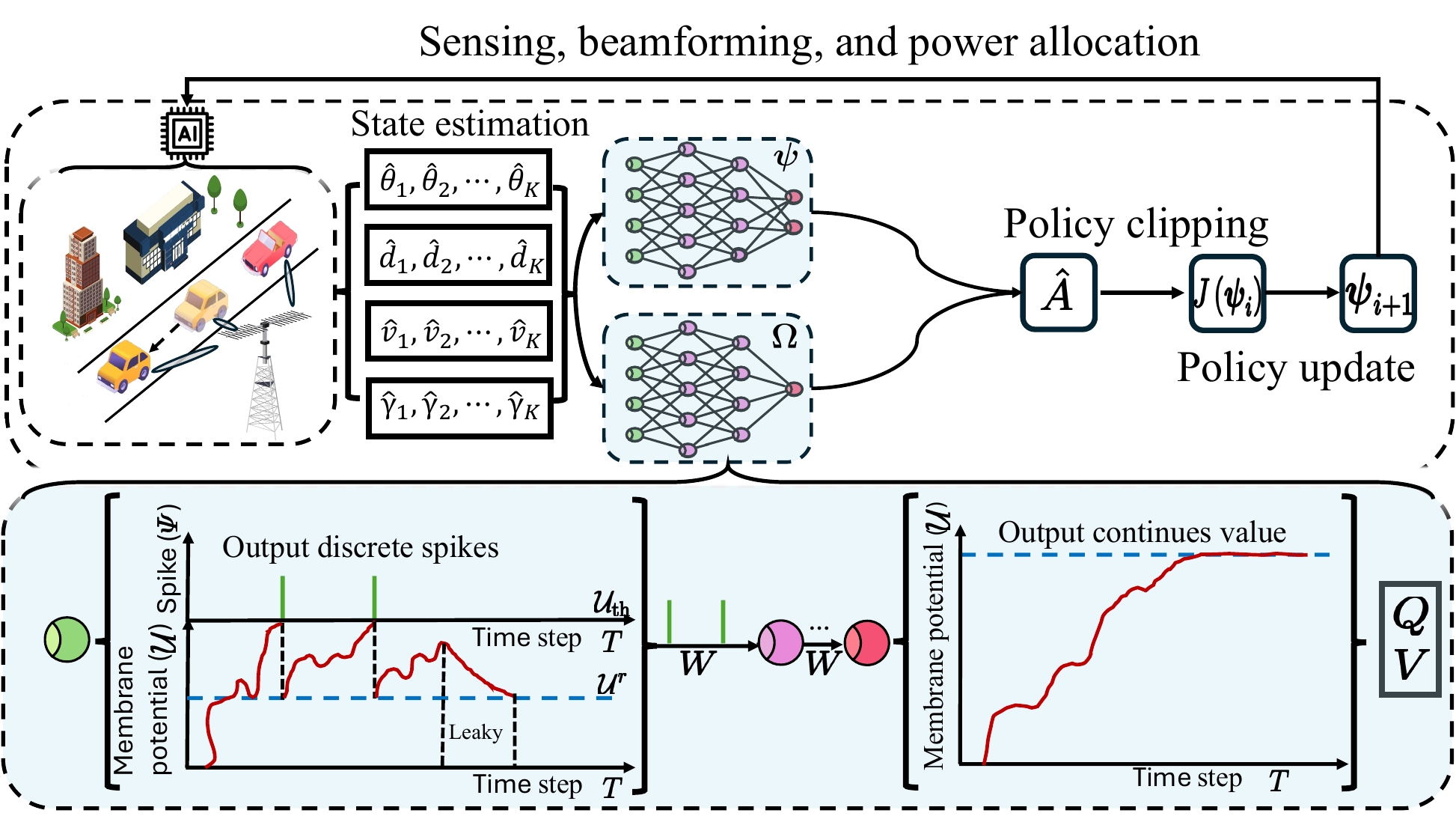}
    \caption{The illustration of the training process for the proposed SNNs-driven Actor-Critic PPO algorithm. The discrete spikes enable this framework to be energy-efficient by dramatically reducing computational complexity.}
    \label{Fig:overall}
    % \vspace{-10pt}
\end{figure}
We utilize the Leaky-Integrate-and-Fire (LIF) mechanism~\cite{yao2022glif} to model spiking neurons. In the LIF-based SNNs framework, spiking neuron-$j$ maintains a time-dependent membrane potential, represented as $\mathcal{U}_j(\tilde{t})$. Note that the $\tilde{t}$ is the time step of SNN with $\tilde{t}\ll n$. The membrane potential evolves over time through two key processes: the integration of incoming inputs, referred to as \textit{charge}, and the gradual decay of the accumulated charge, referred to as \textit{Leaky}, which mimics the behavior of biological neurons~\cite{yao2022glif}. The update of membrane potential $\mathcal{U}_j(\tilde{t})$ is given by:
\begin{equation}
\mathcal{U} _j(\tilde{t}+1)=(1-\lambda) \mathcal{U} _j(\tilde{t})+\lambda\sum_{i=1}^M{W_{i,j}\mathcal{I} _i\left( \tilde{t} \right)},
\label{LIF}
\end{equation}
where $\lambda$ is the membrane leakage coefficient that denotes the rate of charge decay over time, $M$ is the set of neurons connected to neuron $j$, $W_{i,j}$ and $\mathcal{I} _i\left( \tilde{t} \right) $ represent the synaptic weight and input potential from presynaptic neurons, respectively. Once the membrane potential $\mathcal{U} _j\left( \tilde{t} \right) $ exceeds a firing threshold $\mathcal{U}_{\text{th}}$, the neuron releases a spike, represented as $\varPsi _j$. Simultaneously, the membrae potential is reset to $\mathcal{U} ^r\left( \tilde{t} \right) \left( \mathcal{U} ^r\left( \tilde{t} \right) <\mathcal{U}_{\text{th}} \right) $. This dynamic firing mechanism can be expressed as:
\begin{equation}
\begin{cases}
	\varPsi _j=1 ~and~\,\,\mathcal{U} _j(\tilde{t}+1)=\mathcal{U} ^r\left( \tilde{t} \right) \,\,&		\mathrm{if}~\mathcal{U} _j(\tilde{t})\ge \mathcal{U} _{\text{th}}\\
	\varPsi _j=0 ~and~\,\,\mathcal{U} _j(\tilde{t}+1)=\mathcal{U} _j(\tilde{t})&		\mathrm{otherwise}\\
\end{cases}.
\label{spike}
\end{equation}

The details of the LIF neurons in the SNN are illustrated at the bottom of Fig.~\ref{Fig:overall}. In particular, the input is directly encoded and fed into the SNN. This approach preserves fine-grained input information and simplifies preprocessing~\cite{9747906}. The LIF neurons then integrate these inputs over time, enabling the network to capture temporal dependencies through their intrinsic membrane dynamics. When the membrane potential exceeds the firing threshold $\mathcal{U}_{\text{th}}$, the neuron generates discrete spike outputs (0 or 1), which are transmitted to connected neurons in subsequent layers through synaptic weights $W$. Specifically, we remove the membrane potential threshold in the last layer of the SNN. This adjustment allows the last layer to maintain and propagate continuous membrane potentials, enabling the network to generate actions and $V$ tailored to our continuous action space environment, while retaining the spiking behavior in the earlier layers.

\subsubsection{Training of enhanced-DRL algorithm}
Neural networks improve their accuracy by iteratively refining their internal parameters, e.g., weights and biases. This refinement process relies on the differentiability of the network's loss function, which is essential for effective training through back-propagation. In particular, the back-propagation process of SNN-driven DRL can be expressed as:
\begin{equation}
\begin{aligned}
	&\frac{\partial J}{\partial W_{i,j}}=\sum_{\tilde{t}=1}^{\tilde{T}}{\frac{\partial J}{\partial \mathcal{U} _j\left( \tilde{t} \right)}\frac{\partial \mathcal{U} _j\left( \tilde{t} \right)}{\partial W_{i,j}}}\\
	&=\sum_{\tilde{t}=1}^{\tilde{T}-1}{\frac{\partial J}{\partial \mathcal{U} _j\left( \tilde{t} \right)}\frac{\partial \mathcal{U} _j\left( \tilde{t} \right)}{\partial W_{i,j}}} +\frac{\partial J}{\partial \mathcal{U} _j( \tilde{T} )}\frac{\partial \mathcal{U} _j( \tilde{T} )}{\partial W_{i,j}}\\
	&=\sum_{\tilde{t}=1}^{\tilde{T}-1}{\left( \frac{\partial J}{\partial \varPsi _j\left( \tilde{t} \right)}\frac{\partial \varPsi _j\left( \tilde{t} \right)}{\partial \mathcal{U} _j(\tilde{t})}+\frac{\partial J}{\partial \mathcal{U} _j(\tilde{t}+1)}\frac{\partial \mathcal{U} _j(\tilde{t}+1)}{\partial \mathcal{U} _j(\tilde{t})} \right)} \\
	&~\quad \times \frac{\partial \mathcal{U} _j(\tilde{t})}{\partial W_{i,j}}+\frac{\partial J}{\partial \varPsi _j( \tilde{T} )}\frac{\partial \varPsi _j( \tilde{T} )}{\partial \mathcal{U} _j(\tilde{t})}\frac{\partial \mathcal{U} _j(\tilde{t})}{\partial W_{i,j}}\\
\end{aligned},
\label{eq.bp}
\end{equation}
where $\tilde{T}$ denotes the total number of time steps in the SNNs. Specifically, due to the temporal dependencies inherent in SNNs, it is crucial to compute gradients across time steps (i.e., $\tilde{T}$) to capture the dynamics of spiking activity over time, thereby enabling effective learning in time-dependent environments.

It can be found that (\ref{eq.bp}) contains $\frac{\partial J}{\partial \varPsi _j\left( \tilde{t} \right)}$ and $\frac{\partial \varPsi _j\left( \tilde{t} \right)}{\partial \mathcal{U} _j(\tilde{t})}$. In particular, the derivative of $\varPsi _j$ is impulse function (i.e., Dirac Delta function~\cite{hassani2009dirac}), which is given by:
\begin{equation}
\varPsi ^{\prime}\left( x \right) =\left\{ \begin{array}{c}
	+\infty ,x=0\\
	0,~x\ne 0\\
\end{array} \right..
\label{eq: gra1}
\end{equation}
As observed, this function is non-differentiable, which prevents the neural network from being trained normally. To address this issue, we introduce an approximate derivative function as follows~\cite{10636728,fang2021incorporating}:
\begin{equation}
\varphi \left( \varPsi \right) =\frac{1}{\pi}\mathrm{arc}\tan \left( \frac{\pi \eta}{2}\varPsi \right) +\frac{1}{2},
\label{ada}
\end{equation}
where $\varphi \left( \varPsi \right) $ denotes the approximate derivative function that used to approximate the spike output function $\varPsi \left( \mathcal{U} -\mathcal{U} _{\text{th}} \right) $ defined in (\ref{spike}). This approximation ensures that $\varphi \left( \mathcal{U} -\mathcal{U} _{\text{th}} \right) $ is well-defined and has a very close value as $\varPsi \left( \mathcal{U} -\mathcal{U} _{\text{th}} \right) $ even when $\,\,\varPsi =0$. Besides, $\eta > 0$ is the custom parameter. Then, the derivative of $\varphi \left( \varPsi \right)$ is given by:
\begin{equation}
\varphi ^{\prime}\left( \varPsi \right) =\frac{\eta}{2}\times \frac{1}{1+\left( \frac{\pi \eta}{2}\varPsi \right) ^2}.
\label{ada1}
\end{equation}
It can be observed that (\ref{ada}) and (\ref{ada1}) enable SNNs to obtain meaningful gradients while executing the gradient descent algorithm during back-propagation processes. Therefore, \(\varPsi\) can be approximated as \(\underset{x \to 0}{\lim} \varPsi(x) \approx \underset{x \to 0}{\lim} \varphi(x)\). Consequently, the non-differentiable terms \(\frac{\partial J}{\partial \varPsi_j(\tilde{t})}\) and \(\frac{\partial \varPsi_j(\tilde{t})}{\partial \mathcal{U}_j(\tilde{t})}\) in (\ref{eq.bp}) can be replaced with the approximate derivative function $\varphi \left( \cdot \right) $, enabling SNNs to be trained effectively. Note that the approximate derivative function is only used for back-propagation progress, the forward propagation still follows (\ref{LIF}) and (\ref{spike}).

Building on above foundation, we are able to develop an enhanced DRL algorithm that integrates the advanced actor-critic framework with energy-efficient and temporal-dependency SNNs. The details of the training process for the proposed algorithm are illustrated in Fig.~\ref{Fig:overall} and Algorithm~\ref{algo:ppo}. Specifically, the parameters of the SNNs are first initialized randomly (lines 2–4 in Algorithm~\ref{algo:ppo}). At each training iteration $i$, the algorithm collects a set of trajectories $\mathcal{B}_i$ by running the current policy $\boldsymbol{\psi}$ in the considered environment (line 6). After obtaining the cumulative reward $\hat{R}_n$ and advantage function (lines 7-8), the training objective function and the critic network's loss, $\mathcal{C}(\mathbf{\Omega})$, are calculated using (\ref{advanf}) and:
\begin{equation}
\mathcal{C}(\mathbf{\Omega}) = \left(V_n(s_n; \mathbf{\Omega}) - \hat{R}_n\right)^2,
\label{eq:loss of critic}   
\end{equation}
respectively (lines 9–10). With all the obtained objectives and loss functions, the gradients of the actor network, denoted as $\nabla J(\boldsymbol{\psi})$, and then the gradients of critic network, denoted as $\nabla \mathcal{C}(\mathbf{\Omega})$, are computed (lines 10–13). Finally, the networks' parameters are iteratively updated with learning rate $\alpha _{\boldsymbol{a}}$ and $\alpha _{\boldsymbol{c}}$ (lines 14–15), respectively, until the cumulative reward converges to a stationary value.

\begin{algorithm}[h]
%\setstretch{0.7}
\caption{Proposed learning algorithm}
\label{algo:ppo}  % 在这里添加标签
%\SetAlgoLined
%\KwResult{}
\textbf{Input}: \\
Initialize parameter vector $\boldsymbol{\psi }_0$ for actor network, \\ 
Initialize parameter vector $\mathbf{\Omega }_0$ for critic network,  \\
Initialize membrane potential $\mathcal{U}$ for all neurons.\\
\For{$i = 0, 1, 2, \ldots$}{
  Collect set of trajectories $\mathcal{B} _i=\left\{ s_n,a_n,r_n \right\} $ by running policy $\boldsymbol{\psi }_i$ in the environment \\
  Compute cumulative reward $\hat{R}_n=\sum_{n=0}^N{G^nr_n}$ \\
  Compute advantage function $\hat{A}_{n}$ as in (\ref{advanf}) \\
  Compute $J\left( \boldsymbol{\psi } \right) $ as in (\ref{eq:loss of actor})\\
  Compute the loss of critic network $\mathcal{C} \left( \mathbf{\Omega } \right) $ as in (\ref{eq:loss of critic})\\
  Calculate $\nabla J\left( \boldsymbol{\psi } \right) $ and $\nabla \mathcal{C} \left( \mathbf{\Omega } \right) $ as in (\ref{eq.bp}) and (\ref{ada1})\\
  Update the actor network as follows: 
   \begin{equation}
\boldsymbol{\psi }_{i+1}=\boldsymbol{\psi }_i+\alpha _{\boldsymbol{a}}\nabla J\left( \boldsymbol{\psi } \right) 
\label{eq:update of actor}
 \end{equation}\\
  Update the critic network as follows:
  \begin{equation}
\mathbf{\Omega }_{i+1}=\mathbf{\Omega }_i-\alpha _c\nabla \mathcal{C} \left( \mathbf{\Omega } \right) 
\label{eq:loss of crtic}
 \end{equation}
 }
% Initialize membrane potential $\mathcal{U}$ for all neurons\\
 \textbf{Outputs}: $\pi ^*=\mathbb{P} (a_n|s_n;\boldsymbol{\psi })$
\end{algorithm}

\section{Performance Evaluations} \label{PE}
\begin{figure*}[t]
	\centering
	\begin{subfigure}[b]{0.26\linewidth}
		\centering
		\includegraphics[width=1.0\linewidth]{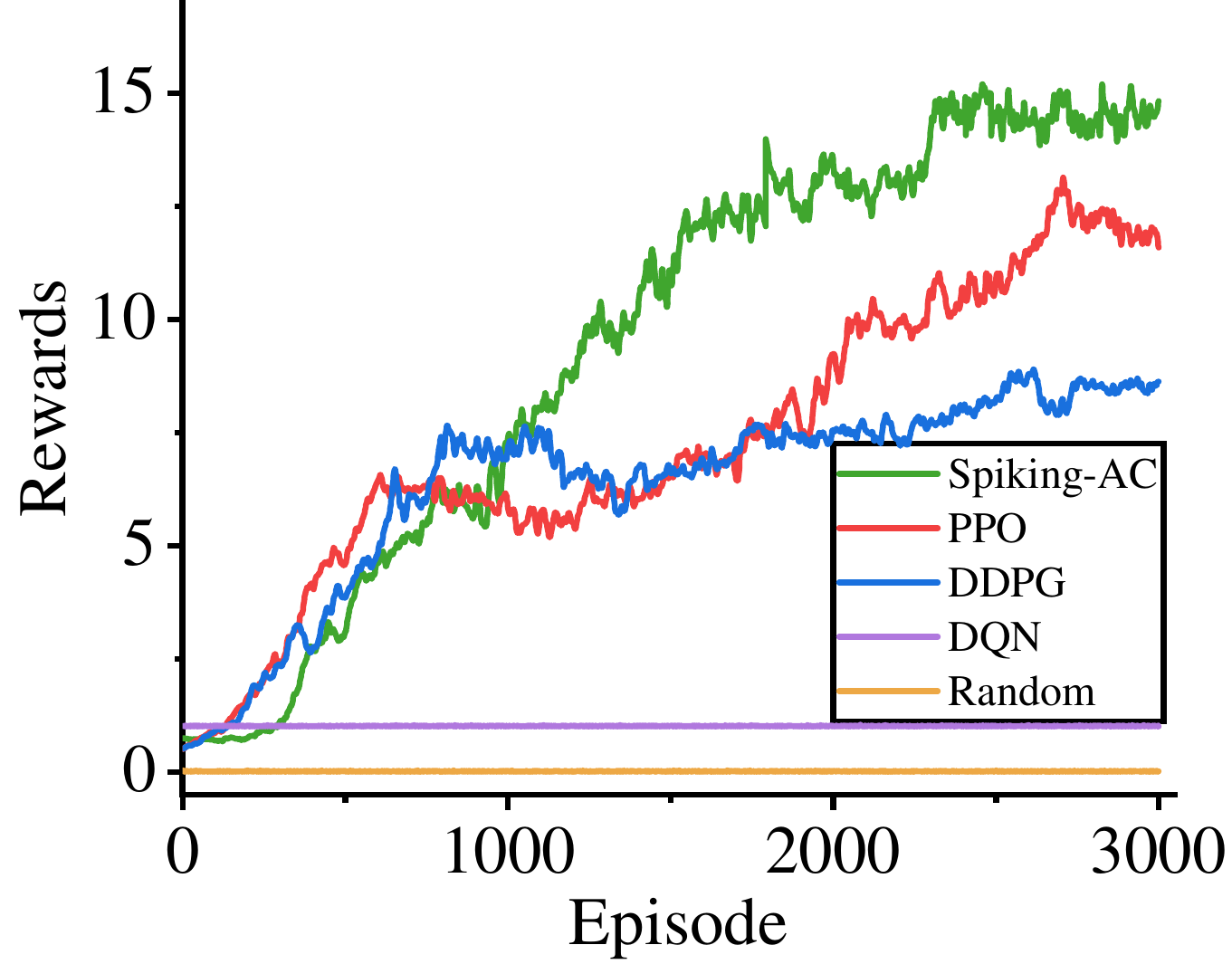}
		\caption{}
	\end{subfigure}%
	% \end{subfigure}%
        ~
	\begin{subfigure}[b]{0.26\linewidth}
		\centering
		\includegraphics[width=1.0\linewidth]{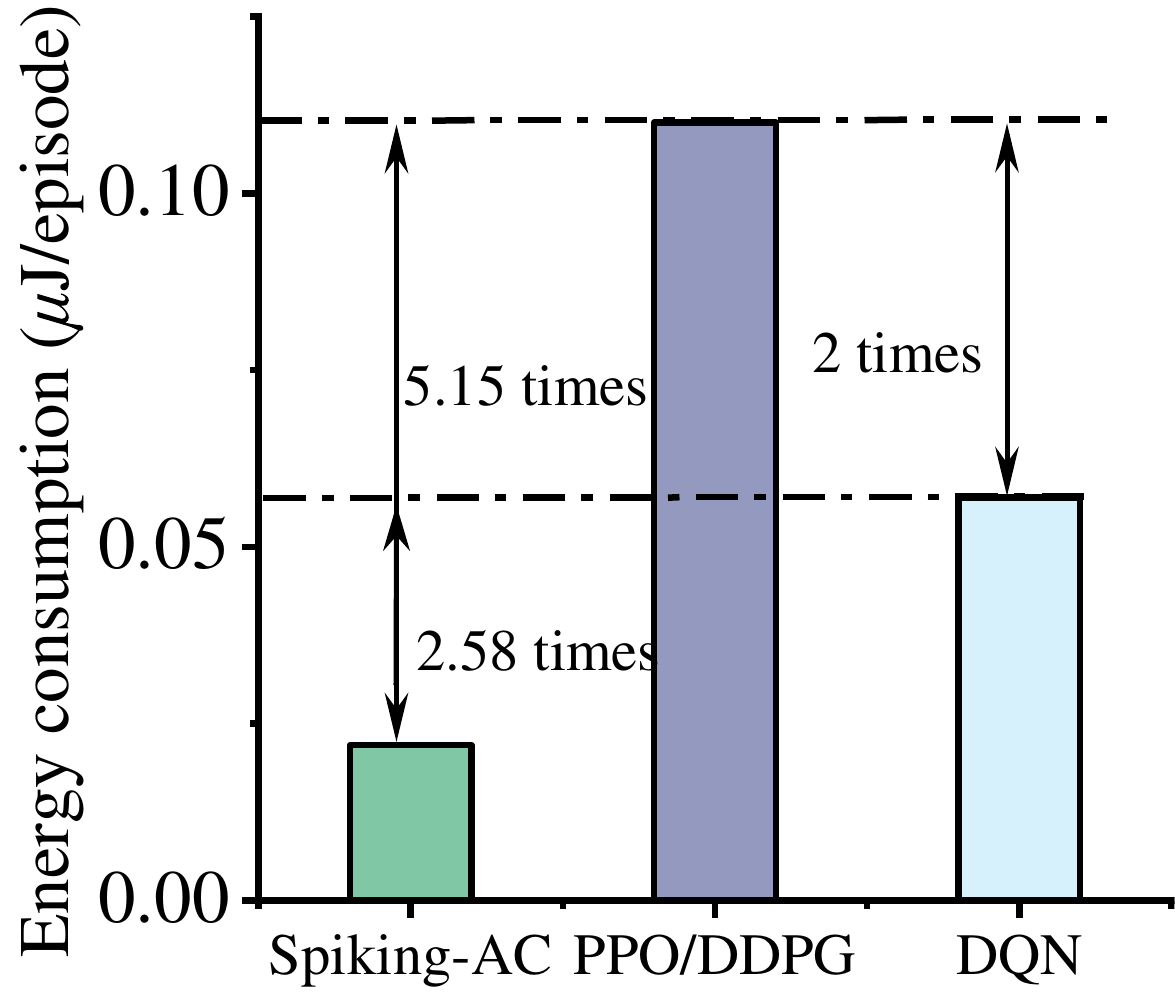}
		\caption{}
	\end{subfigure}%
	~
	\begin{subfigure}[b]{0.26\linewidth}
		\centering
		\includegraphics[width=1.0\linewidth]{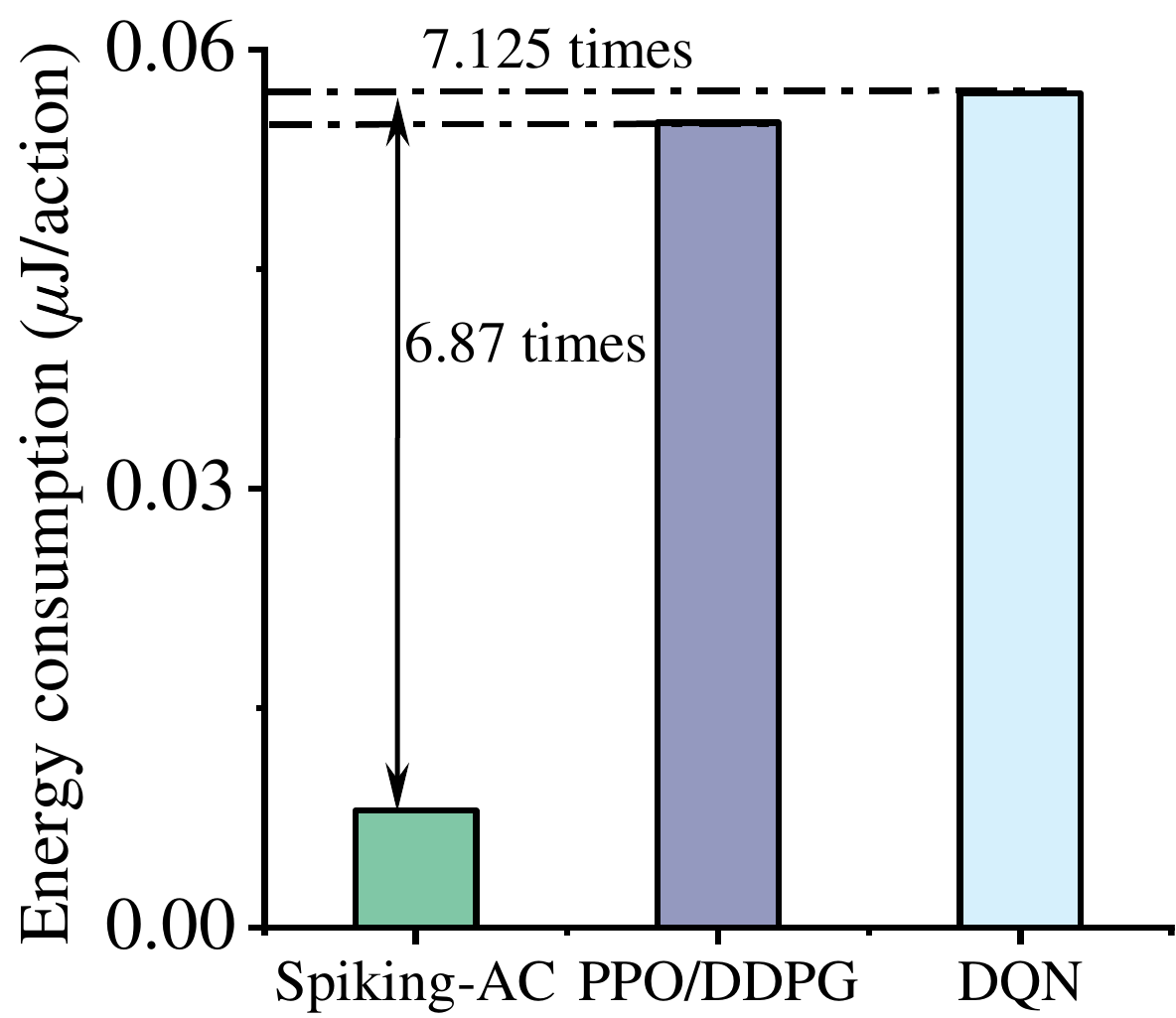}
		\caption{}
	\end{subfigure}
	\caption{(a) Comparison of the rewards in training process, (b) Comparison of energy consumption in training process, and (c) Comparison of energy consumption in inference process.}
	\label{fig:training-results}
    % \vspace{-10pt}
\end{figure*}
\subsection{Simulation Settings}\label{simulation setting}
\textit{1) V2X Network}: We consider a scenario with three vehicles (i.e., $K=3$), each equipped with a single antenna, driving along a straight road. The vehicles enter the coverage area of an RSU with $N_{\text{TA}} = N_{\text{RA}} = 32$ antennas, operating at a carrier frequency $f_c = 30$~GHz~\cite{liu2020radar}.  Without loss of generality, the coordinate of the RSU is set as (0, 0) and the initial positions of the vehicles are (-5, 10), (-15, 10), and (-25, 10), respectively. For the state evolution model of the vehicles, we consider that their average velocities follow $v \sim \mathrm{Unif}(10~\text{m/s}, 14~\text{m/s})$, i.e. approximately 36-50.4 km/h. The evolution noise parameters are set as $\sigma_{\theta} = 0.02^{\circ}$, $\sigma_d = 0.2~\text{m}$, and $\sigma_v = 0.5~\text{m/s}$~\cite{9246715}. The time duration is $\Delta T = 0.02$~s, and the total driving time instant for the vehicles is $N = 100$. Additionally, the $\kappa=10+10j$ and $\xi=10$~\cite{liu2020radar}. The noise power for both sensing and communication signals is $\sigma_z^2 = \sigma_c^2 = -80$~\text{dBm}. For the measurement noise parameters, we set $\alpha_{\tau} = 1 \times 10^{-9}$ and $\alpha_{\mu} = 2 \times 10^3$~\cite{9246715}.

\textit{2) Algorithm Parameter Setups}: We utilize a three-layer fully connected neural network to implement the DRL framework, consisting of an input layer, a hidden layer, and an output layer. This shallow network architecture offers sufficient representational capacity for beamforming task while enabling lower energy consumption.
The number of neurons in the input and output layers corresponds to the dimensions of the state and action spaces, respectively, while the hidden layer contains 128 neurons.
The batch size and discount factor are set to \{512, 0.99\}. Besides, the policy-clipping factor is configured with $\epsilon = 0.2$~\cite{10742922,ppo}. Moreover, the learning rates for the actor and critic networks are set to $5\times 10^{-5}$ and $5\times 10^{-4}$, respectively. For the SNNs, we set the parameters of the LIF-based neurons as $\tilde{T} = 6$, $\eta = 3$, $\mathcal{U}_{\text{th}} = 1$, $\mathcal{U}^r = 0$, and $\lambda = \frac{1}{2}$~\cite{10636728}. 

\textit{3) Baselines}: We conduct performance comparisons between our proposed method, denoted as Spiking Actor-Critic (Spiking-AC), and several state-of-the-art algorithms, including the following: Proximal Policy Optimization (PPO)~\cite{schulman2017proximal}, Deep Deterministic Policy Gradient (DDPG)~\cite{lillicrap2015continuous}, DQN~\cite{mnih2015human}, and Random scheme. 
% \vspace{-10pt}
\subsection{Simulation Results}
\subsubsection{Convergence Performance and Energy Consumption}
We first illustrate the convergence performance of the proposed algorithm and the baselines in Fig.~\ref{fig:training-results}(a). In Fig.~\ref{fig:training-results}(a) we can observe that Spiking-AC and PPO demonstrate the fastest convergence and achieve the highest rewards, highlighting their ability to effectively balance the dual objectives of maximizing the sum-rate and enhancing sensing accuracy in complex and dynamic V2X scenarios. Conversely, the DDPG algorithm shows moderate performance, with slower reward accumulation and reduced optimization efficiency, indicating it is less efficient than Spiking-AC and PPO in optimizing beamforming and power allocation schemes in the high-dimensional and uncertain action and state spaces considered. Specifically, DDPG relies on an off-policy learning approach with a replay memory buffer to store and sample past experiences~\cite{lillicrap2015continuous}. This off-policy learning mechanism occasionally introduces suboptimal samples that hinder training efficiency while consuming significant computational resources~\cite{10486201}. In contrast, the proposed Spiking-AC and PPO algorithms utilize an on-policy learning approach and employ a policy-clipping technique as in (\ref{eq:policy-clipping}) that enables network parameter updates directly from interactions with the environment, reducing the need for large replay memory buffers typically required by off-policy methods. This design allows Spiking-AC to achieve more efficient and streamlined learning while significantly reducing computational resource consumption. Additionally, compared to PPO, Spiking-AC further benefits from its spiking-based neural network architecture, which not only improves convergence and achievable rewards but also significantly reduces energy consumption during the training and inference processes, which is detailed in the following. 

Furthermore, the DQN algorithm and the Random scheme are unable to to find improved solutions during the learning process, with the the Random approach shows the poorest performance. Interestingly, despite being a learning-based algorithm, DQN demonstrates the poor performance in the uncertain and dynamic scenario characterized by continuous state and action spaces. This is primarily because DQN is inherently designed for discrete action spaces. Although discretizing the continuous action space into a finite set of representative actions partially addresses this limitation, it significantly compromises precision and flexibility. Additionally, storing these discretized actions in the replay memory buffer further hinders exploration, severely restricting the algorithm's adaptability to dynamic environments. As expected, the Random approach shows the poorest performance

We then evaluate the energy consumption during model training and inference. In particular, the total number of floating-point operations (FLOPs) is used to quantify the computational workload during training and inference. These operations are then mapped to energy consumption by assigning hardware-specific energy coefficients to different operation types (i.e., MAC or AC)~\cite{panda2020toward,10636728}. As discussed in Section~\ref{algorithm}, the Spiking-AC algorithm leverages binary events (i.e., spikes) for computation, requiring only AC operations to execute the dot product without the need for multipliers. In contrast, baseline algorithms rely on traditional MAC operations. For a three-layers network, the FLOPs for Spiking-AC and the baseline algorithms can be calculated as: $FLOPS_{spiking}=\sum_{i=1}^2{\left( N_i\times M_i\times \varsigma _i \right)}$ and $FLOPS_{baselines}=\sum_{i=1}^3{\left( N_i\times M_i \right)}$, respectively, where $N_i$ and $M_i$ represent the input and output dimensions of the $i$-th fully connected layer, $\varsigma_i$ denotes the spike firing rate~\cite{panda2020toward,10636728}. Note that the final layer of this fully connected network does not use the LIF model, as the actor and critic networks require continuous-valued outputs.   
Accordingly, the energy consumption of Spiking-AC and the baseline algorithms can be expressed as: $E_{spiking}=\left( E_{AC}\times FLOPS_{spiking}\times \tilde{T} \right) +E_{MAC}\times N_3\times M_3$ and $E_{baseline}=E_{MAC}\times FLOPS_{baselines}$, respectively, where $E_{AC}=0.1$pJ and $E_{MAC}=3.2$pJ are the energy consumption per operation for AC and MAC~\cite{han2015learning}, respectively. 

We show the comparison of average energy consumption per training and inference episode (i.e., the energy consumed by the RSU for executing a single action using the trained model) in Fig.~\ref{fig:training-results}(b) and Fig.~\ref{fig:training-results}(c). Notably, the energy consumption of the PPO and DDPG algorithms is identical because they share the same actor-critic framework\footnotemark\footnotetext{The off-policy algorithms such as DDPG consume more energy than on-policy algorithms due to the overhead of updating the replay memory buffer. However, in this work, we focus exclusively on the energy consumption of the neural network.}. Additionally, the energy consumption of PPO and DDPG is approximately twice that of DQN, as DQN uses only a single network. As shown in Fig.~\ref{fig:training-results}(b) and (c), Spiking-AC demonstrates superior efficiency, consuming 5.15 times less energy than PPO and DDPG during training and 2.58 times less than DQN. Notably, during inference, Spiking-AC further reduces energy consumption due to its spiking-based architecture.

Based on above analysis, we can conclude that Spiking-AC achieves robust and efficient training and inference performance while significantly conserving energy. Compared to conventional methods, it reduces energy consumption by over 50\% while delivering superior performance, underscoring its practicality for energy-constrained V2X networks.

\subsubsection{Communication Performance}
\begin{figure}[t]
	\centering
	\begin{subfigure}[b]{0.5\linewidth}
		\centering
		\includegraphics[width=1.0\linewidth]{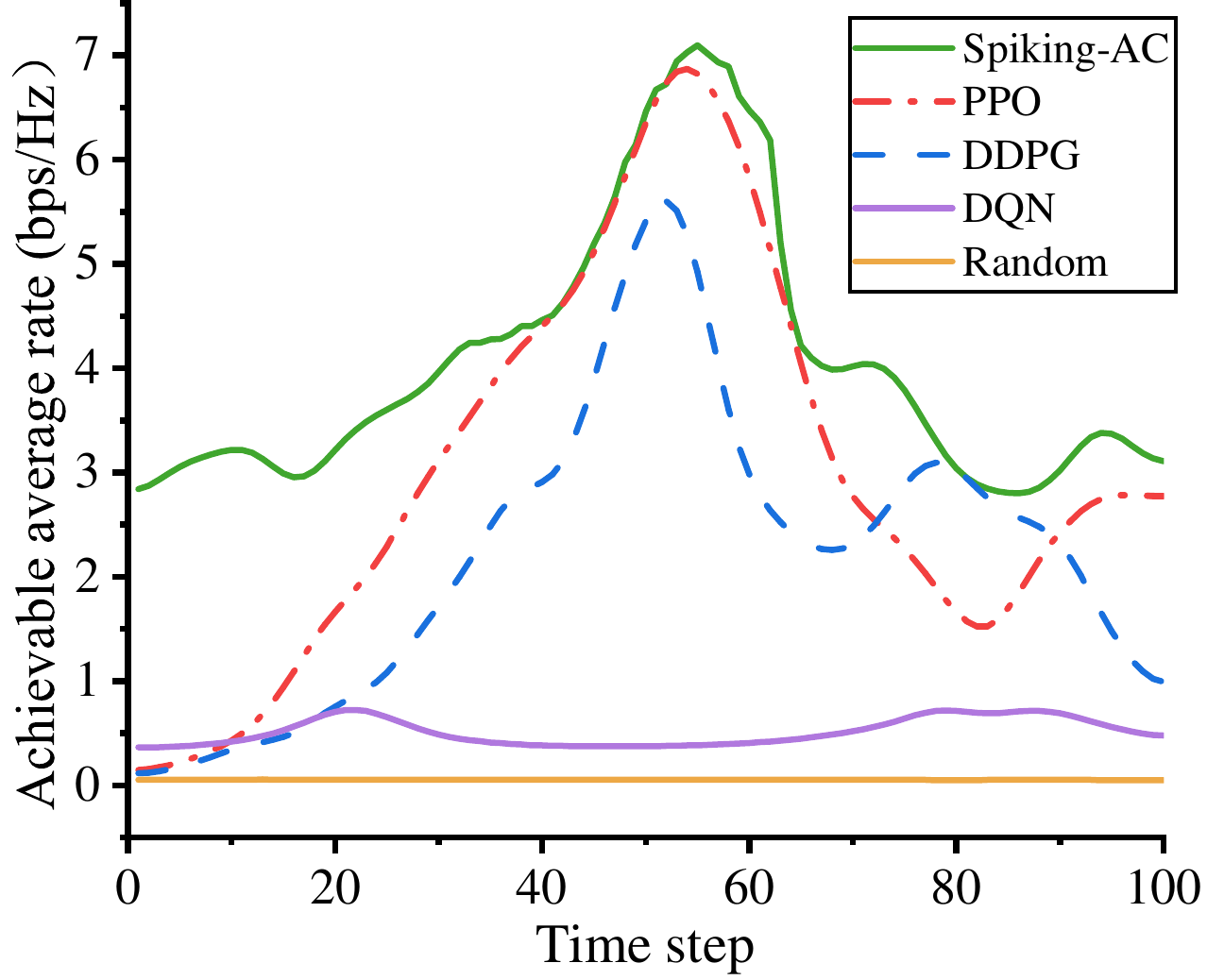}
		\caption{}
	\end{subfigure}%
	\begin{subfigure}[b]{0.5\linewidth}
		\centering
		\includegraphics[width=1.0\linewidth]{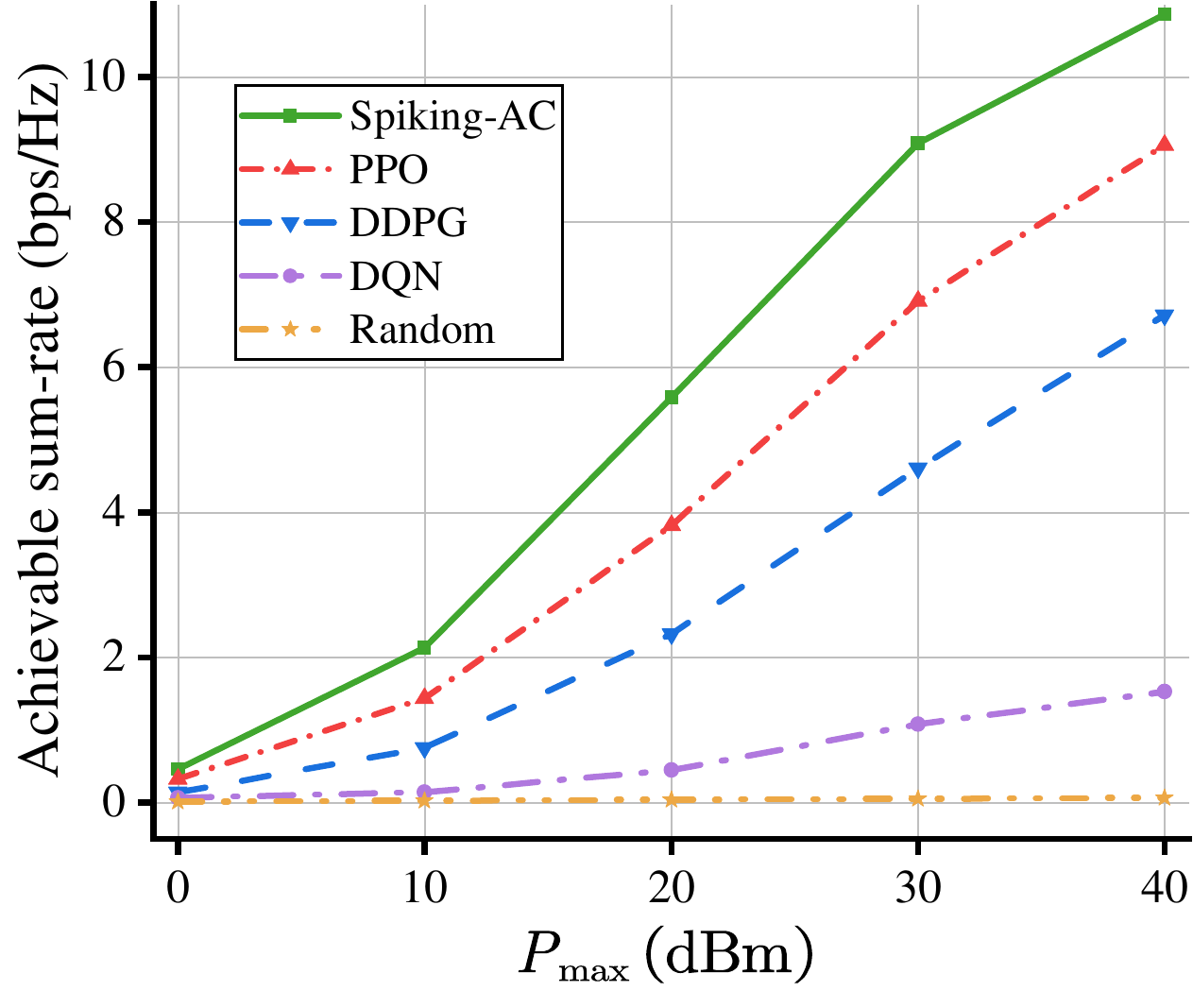}
		\caption{}
	\end{subfigure}
	\caption{(a) The achievable average sum-rate at $P_{\max}$ = 40 dBm and (b) The achievable sum-rate vs. maximum transmit power.} 
	\label{fig:communicationrate}
    % \vspace{-17pt}
\end{figure}
We then present the achievable communication rates during the vehicles’ driving progress. 
As shown in Fig.~\ref{fig:communicationrate}(a), the proposed Spiking-AC algorithm achieves the highest transmit rate during the first 40 time steps, outperforming other algorithms. This highlights Spiking-AC algorithm's ability to maintain robust communication rates even for vehicles at long distances, enabled by its efficient beamforming and power allocation schemes driven by the SNNs and policy-clipping technique. By jointly optimizing beamforming and transmission power, Spiking-AC effectively mitigates path loss, outperforming other algorithms that unable to optimize these parameters simultaneously. 

The PPO baseline, which employs the same policy-clipping technique as Spiking-AC, achieves the second-best performance during this period. However, Spiking-AC surpasses PPO by integrating SNNs, which enhance training efficiency and inference robustness. DDPG, however, despite sharing the actor-critic framework with Spiking-AC and PPO, is unable to achieve comparable communication rates, primarily due to the absence of the policy-clipping technique that is crucial for training stability and inference efficiency. Over time, the communication rates increase as vehicles approach the RSU, peaking when they are closest (i.e., 40-60 time steps) and subsequently declining as they move away. Interestingly, the DQN demonstrates contrasting results. This is because, as vehicles approach the RSU, the challenges of managing interference, optimizing resource allocation, and adapting to the dynamic environment become more pronounced. Due to its lack of coordination and adaptability, the DQN scheme becomes increasingly ineffective in these conditions, resulting in worse performance compared to when vehicles are farther from the RSU. This underscores the critical importance of intelligent resource management in multi-vehicle scenarios.

Next, we vary the maximum transmit power at the RSU, i.e., $P_{\max}$, to evaluate the impact of power allocation on communication performance. As shown in Fig.~\ref{fig:communicationrate}(b), the proposed Spiking-AC algorithm consistently achieves the highest sum-rate across all transmit power levels, outperforming baseline algorithms such as PPO, DDPG, DQN, and Random allocation. Notably, at a maximum transmit power of 40 dBm ($P_{\max}$ = 40 dBm), Spiking-AC achieves an sum-rate of 10.864 bps/Hz, surpassing PPO (9.058 bps/Hz) and DDPG (6.716 bps/Hz) by 19.96$\%$ and 61.7$\%$, respectively. Furthermore, as the maximum transmit power of the RSU decreases, the performance gap among the algorithms diminishes. However, Spiking-AC retains its superiority by efficiently optimizing beamforming and power allocation strategies, delivering consistently robust performance even under stringent power constraints.

\subsubsection{Sensing Performance}
\begin{figure}[t]
	\centering
	\begin{subfigure}[b]{0.5\linewidth}
		\centering
		\includegraphics[width=1.0\linewidth]{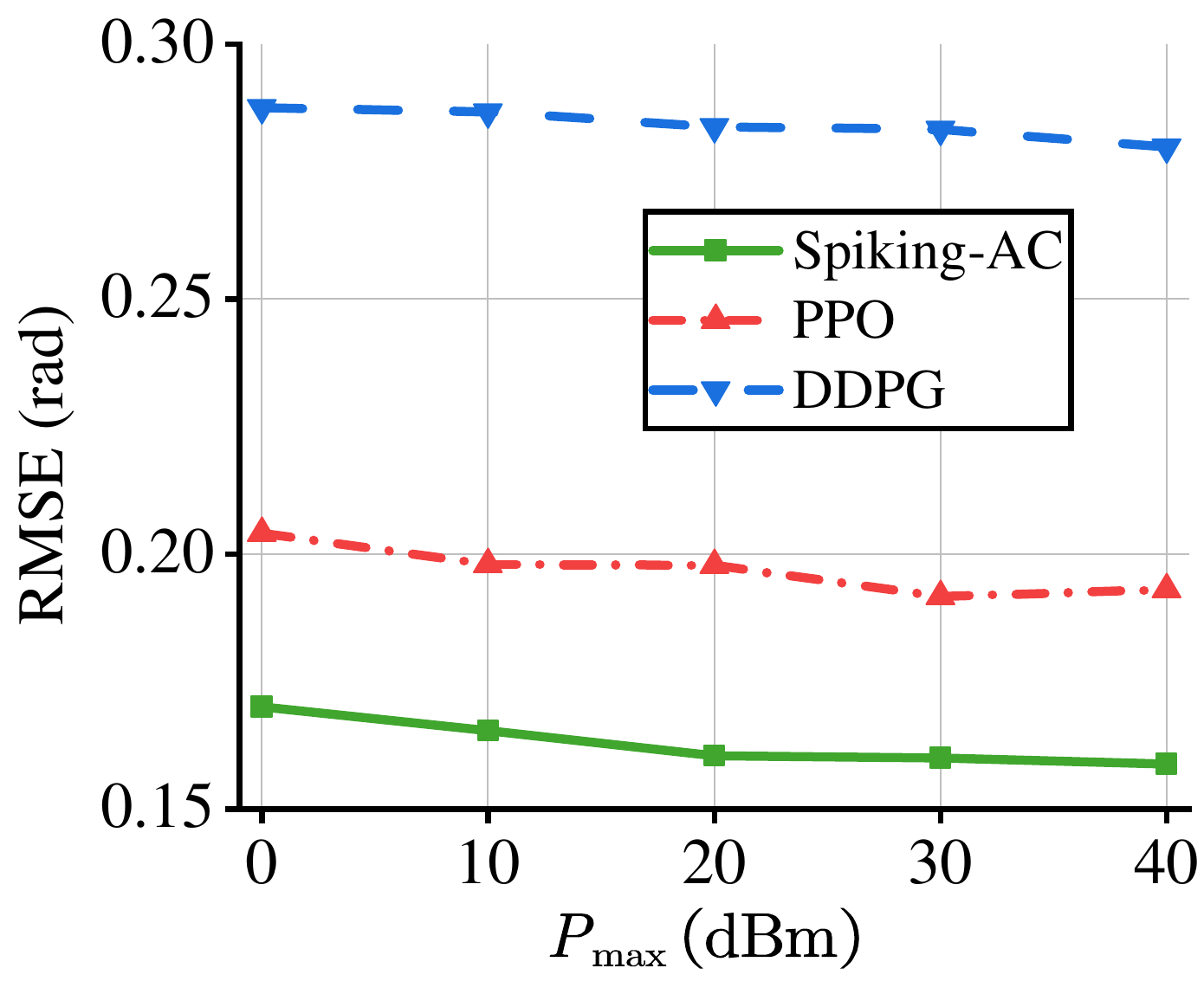}
		\caption{}
	\end{subfigure}%
	\begin{subfigure}[b]{0.5\linewidth}
		\centering
		\includegraphics[width=1.0\linewidth]{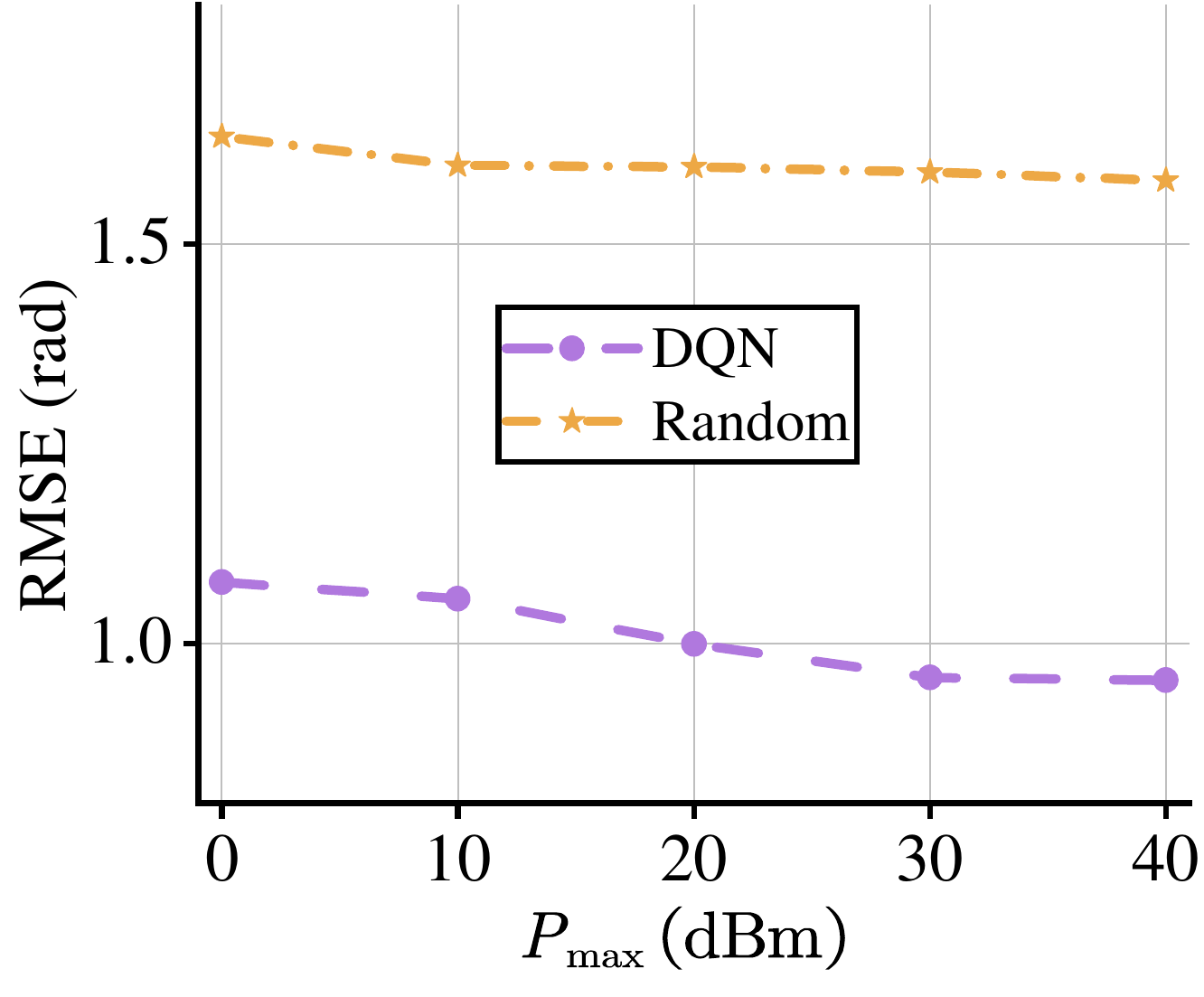}
		\caption{}
	\end{subfigure}
	\caption{Comparison of sensing performance for angle $\theta$.} 
	\label{fig:sensing theta}
    % \vspace{-10pt}
\end{figure}
We evaluate the sensing performance by analyzing the RMSE values achieved by different algorithms. It is worth noting that the sensing accuracy exhibits a trend similar to the communication sum-rate shown in Fig.~\ref{fig:communicationrate}(a), peaking when the vehicles are closest to the RSU. To ensure a comprehensive comparison, the RMSE values are averaged over the entire process.

As shown in Fig.~\ref{fig:sensing theta}, the sensing performance for $\theta$ improves across all algorithms as the transmit power increases. This is because higher transmit power improves the RSU’s ability to process the echoes of sensing signals, as defined in (\ref{eq:crlb-theta}), thereby reducing noise impact and enabling more accurate angle estimation. Among the algorithms, Spiking-AC consistently achieves the lowest RMSE across all power levels, improving from 0.1654 rad at 0 dBm to 0.1588 rad at 40 dBm. Compared to PPO, Spiking-AC achieves an RMSE reduction of 1.31\% at 10 dBm and 1.46\% at 40 dBm, and significantly outperforms DDPG (7.53\% reduction), DQN (83.35\% reduction), and Random (89.95\% reduction) in optimizing beamforming and power allocation strategies for sensing tasks.

\begin{figure}[t]
	\centering
	\begin{subfigure}[b]{0.5\linewidth}
		\centering
		\includegraphics[width=1.0\linewidth]{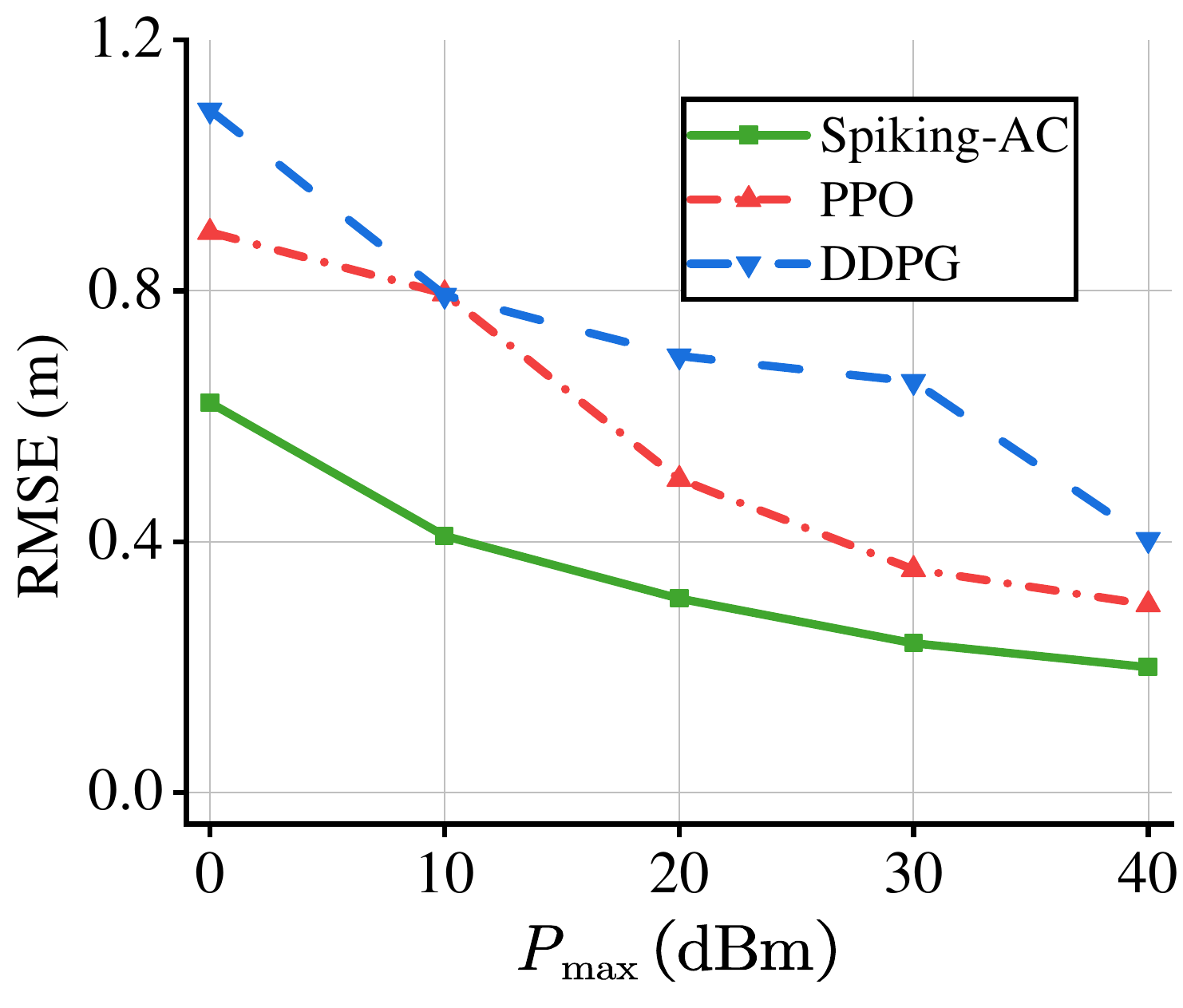}
		\caption{}
	\end{subfigure}%
	\begin{subfigure}[b]{0.5\linewidth}
		\centering
		\includegraphics[width=1.0\linewidth]{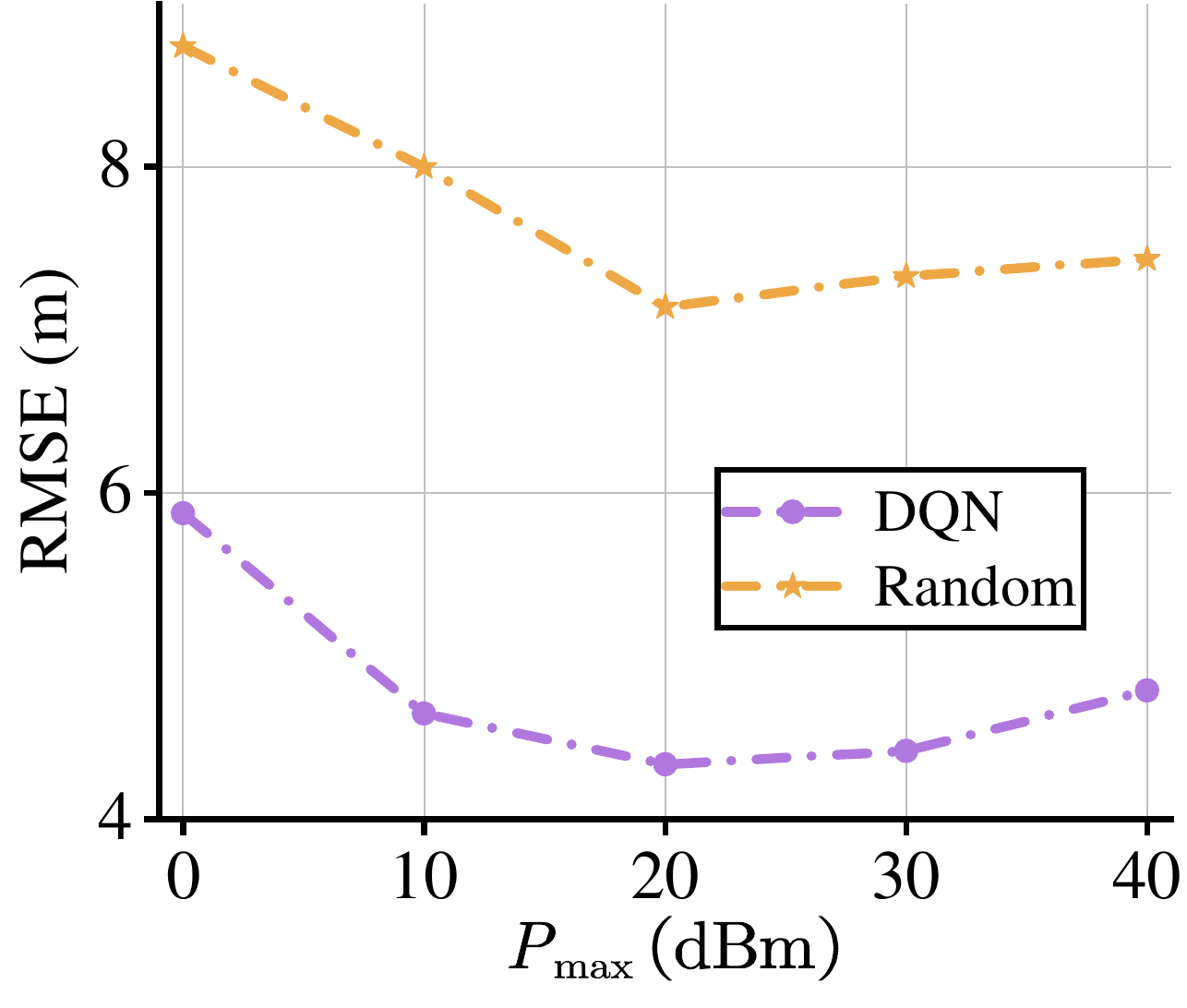}
		\caption{}
	\end{subfigure}
	\caption{Comparison of sensing performance for distance $d$.} 
	\label{fig:sensing distance}
    % \vspace{-10pt}
\end{figure}
In Fig.~\ref{fig:sensing distance}, the sensing performance for distance generally improves (i.e., RMSE decreases) with increasing transmit power across most algorithms. Specifically, Spiking-AC achieves the best performance, with RMSE decreasing from 0.62175 at 0 dBm to 0.2 at 40 dBm, highlighting its superior capability to leverage SNN-based policies for optimizing sensing performance under varying power levels.
In comparison, PPO reduces its RMSE from 0.89357 to 0.3, while DDPG decreases from 1.088 to 0.40358, demonstrating their ability to adapt to higher transmit power. However, their performance remains inferior to Spiking-AC. In contrast, DQN and Random schemes display inconsistent trends, exhibiting non-monotonic behavior with RMSE worsening to 4.79 and 7.435 at 40 dBm, respectively. For DQN, this can be attributed to its reliance on a discrete action space, which lacks the granularity required to finely optimize transmit power, resulting in interference and degraded sensing accuracy at higher power levels. Similarly, the Random scheme suffers from high inference errors caused by increased power levels.

At 40 dBm, Spiking-AC outperforms PPO and DDPG by 33.33\% and 50.37\%, respectively. Moreover, Spiking-AC achieves significant RMSE reductions of 95.83\% and 97.31\% compared to DQN and Random, respectively. These results emphasize Spiking-AC’s ability to effectively minimize estimation error and maintain superior sensing performance as transmit power increases. This consistent improvement is attributed to Spiking-AC's advanced framework, which integrates SNNs for processing complex information and employs policy-clipping techniques to ensure stable training and efficient inference. By effectively balancing beamforming and power allocation, Spiking-AC minimizes interference and path loss, thereby maintaining robust sensing performance.

% \subsubsection{Inference Latency}
% \vspace{-10pt}
\section{Conclusion}\label{conclusion}
In this work, we have proposed an energy-efficient and intelligent ISAC system for V2X networks. In particular, we have first utilized an MDP framework to model the complex environmental dynamics, enabling the RSU to develop efficient beamforming schemes based solely on current sensing state information. To enhance system intelligence, we have then introduced a DRL algorithm based on the Actor-Critic framework with a policy-clipping technique, facilitating the joint optimization of beamforming and power allocation strategies to ensure high communication rates and accurate sensing. Furthermore, to address the energy demands of neural networks, we have integrated SNNs into the DRL algorithm, leveraging their discrete spikes and temporal characteristics to significantly reduce energy consumption during training and inference while simultaneously improving system performance. Simulation results have shown that our proposed scheme outperforms baseline methods in terms of energy consumption, communication rates, and sensing accuracy. These findings validate the potential of the proposed system to enable intelligent, robust, and sustainable V2X connectivity in dynamic and resource-constrained environments.

\bibliographystyle{IEEEtran}
\bibliography{bibRef}

\end{document}